\documentclass[a4paper,11pt]{article}
\usepackage{amssymb,graphicx,amsmath,color}

\renewcommand{\theequation}{\thesection.\arabic{equation}}
\input{tcilatex}

\newtheorem{rules}{Rules}[section]
\newtheorem{propo}{Proposition}
\newtheorem{defi}[rules]{Definition}

\newcommand{\halmos}{\rule{1ex}{1.4ex}}
\newcommand{\eproof}{\hspace*{\fill}\mbox{$\halmos$}}

\newcommand{\bc}{\begin{center}}
\newcommand{\ec}{\end{center}}
\def\ba#1{\begin{array}{#1}\displaystyle}
\newcommand{\ea}{\end{array}}

\newcommand{\beq}{\begin{equation}}
\newcommand{\eeq}{\end{equation}}
\newcommand{\beqa}{\begin{eqnarray}}
\newcommand{\eeqa}{\end{eqnarray}}

\newcommand{\n}{\nonumber\\}
\newcommand{\bi}{\begin{itemize}}
\newcommand{\ei}{\end{itemize}}
\def\mato#1{\left(\ba{#1}} % exemple: \mato{cc} a & b \\ c & d \matf
\def\matf{\ea\right)}

\def\lt#1{\left#1}
\def\rt#1{\right#1}

\def\h#1{\hat{#1}}
\def\b#1{\bar{#1}}
\def\frc#1#2{\frac{#1}{#2}}

\newcommand{\bra}{\langle}
\newcommand{\ket}{\rangle}
\newcommand{\Z}{{\mathbb{Z}}}

\newcommand{\C}{{\mathbb{C}}}

\newcommand{\ep}{\epsilon}

\newcommand{\Tr}{{\rm Tr}}

\newcommand{\tw}{{\cal T}}

\newcommand{\aux}{{\text{aux}}}

\begin{document}

\setcounter{page}{0} \topmargin0pt \oddsidemargin0mm \renewcommand{%
\thefootnote}{\fnsymbol{footnote}} \newpage \setcounter{page}{0}
\begin{titlepage}
\vspace{0.2cm}
\begin{center}
{\Large {\bf Permutation operators, entanglement entropy, and the
XXZ spin chain in the limit $\Delta\to-1^+$}}

\vspace{0.8cm} {\large \text{Olalla A.~Castro-Alvaredo$^{\bullet}$
and Benjamin Doyon$^{\circ}$}}

\vspace{0.2cm}
{$^{\bullet}$  Centre for Mathematical Science, City University London, \\
Northampton Square, London EC1V 0HB, UK}\\
{$^{\circ}$  Department of Mathematics, King's College London,
\\
Strand, London WC2R 2LS, UK}
\end{center}
\vspace{1cm} In this paper we develop a new approach to the
investigation of the bi-partite entanglement entropy in integrable
quantum spin chains. Our method employs the well-known replica
trick, thus taking a replica version of the spin chain model as
starting point. At each site $i$ of this new model we construct an
operator $\mathcal{T}_i$ which acts as a cyclic permutation among
the $n$ replicas of the model. Infinite products of $\tw_i$ give
rise to local operators, precursors of branch-point twist fields
of quantum field theory. The entanglement entropy is then
expressed in terms of correlation functions of such operators.
Employing this approach we investigate the von Neumann and R\'enyi
entropies of a particularly interesting quantum state occurring as
a limit (in a compact convergence topology) of the
antiferromagnetic XXZ quantum spin chain. We find that, for large
sizes, the entropy scales logarithmically, but not conformally.

 \vfill{
\hspace*{-9mm}
\begin{tabular}{l}
\rule{6 cm}{0.05 mm}\\
$^\bullet \text{o.castro-alvaredo@city.ac.uk}$\\
$^\circ \text{benjamin.doyon@kcl.ac.uk}$\\
\end{tabular}}

\renewcommand{\thefootnote}{\arabic{footnote}}
\setcounter{footnote}{0}

\end{titlepage}
\newpage
\section{Introduction}
Entanglement is a fundamental property of quantum systems.
Historically, there has been great interest in developing
efficient theoretical measures of entanglement, one of which is
known as the bi-partite entanglement entropy \cite{bennet}. This
entropy measures the amount of quantum
entanglement, in some pure quantum state, between the degrees of
freedom associated to two sets of independent observables whose
union is complete on the Hilbert space.

\begin{figure}[h!]
\begin{center}
\includegraphics[width=10cm,height=2cm]{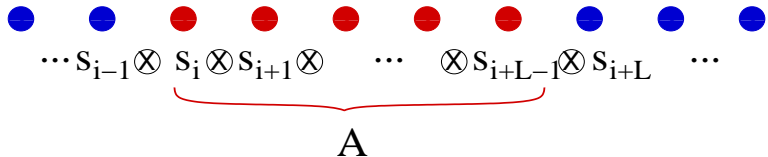}\\
  \caption{A region $A$ of length $L$ of a quantum spin chain}\label{chain}
  \end{center}
\end{figure}

To make this definition more precise, let us consider a
one-dimensional quantum spin chain such as the one depicted in
Fig.~\ref{chain} and let $|{\rm gs}\ket$ be the ground state of
that chain.
We will subdivide the chain in two regions, $A$ and its complement
$\bar{A}$. In the figure, region $A$ is the block of $L$ spins in
red whereas $\bar{A}$ is the rest of the chain, in blue.
Associated to this quantum spin chain there is a Hilbert space
which can be expressed as a tensor product of local Hilbert spaces
associated to its sites. This can be written as a tensor product
of the two Hilbert spaces associated to the regions $A$ and
$\bar{A}$:
\begin{equation} \label{Hdecomp} {\cal H} = {\cal A} \otimes
\b{{\cal A}}.
\end{equation}
The bi-partite entanglement entropy is the von Neumann entropy of
the reduced density matrix $\rho_A$ associated to $A$, that is:
\beq\label{defeegs} S_A = -\Tr_{{\cal A}} \left( \rho_A\log
\rho_A\right) ~,\quad \rho_A = \Tr_{\b{{\cal A}}} \left(|{\rm
gs}\ket \bra {\rm gs}|\right). \eeq It is the entropy of the
region $A$ with respect to the rest of the chain, regarded as its
environment. Another measure of entanglement that is widely
studied in the literature is the R\'enyi entropy \cite{renyi},
given by
\begin{equation}
    S_A^{\text{R\'enyi}}(n)=\frac{\log\left(\Tr_{{\cal A}}\left( \rho_A^n
    \right)\right)}{1-n},\label{renyi}
\end{equation}
whose $n\rightarrow 1$ limit gives the von Neumann entropy.

The bi-partite entanglement entropy of one-dimensional
(integrable) quantum spin chains has been extensively studied in
the literature from different points of view. One may for instance
consider a quantum spin chain divided in two parts, both of which
are infinitely long. If the spin chain has a finite gap, the
entropy saturates to a constant value which depends on the
parameters of the model under consideration and can in some cases
be computed analytically \cite{Weston,ravanini,Ercolessi:2010eb}.
The methods used are based on the algebraic Bethe ansatz or on the
corner transfer matrix. Another interesting set-up are quantum
spin chains which are infinitely long but where the block $A$ is
kept of finite length. In the critical regime, it has been found
through a combination of numerical and analytical approaches (many
of which can only be applied for integrable models)
\cite{Latorre1,Latorre2,Latorre3,Jin,Lambert,KeatingM05} that the
entropy of a block diverges logarithmically with the size of the
block, as expected from conformal field theory (CFT)
considerations
\cite{CallanW94,HolzheyLW94,Calabrese:2004eu,Calabrese:2005in}.

The regions $A$ and $\bar{A}$ can take many different shapes and
need not necessarily consist of connected blocks of spins. That
is, we may also consider a situation in which both $A$ and
$\bar{A}$ are made out of the union of several disjoint blocks of
spins. The entropy of disconnected regions has been the object of
numerical and analytical study recently in the context of CFT
\cite{disco4,casini-dis,Casini:2008wt,disco1,disco12} and quantum
spin chains and lattice models \cite{disco3,disco2}.

The goal of this paper is two-fold. First, we introduce
the quantum chain precursors of
branch-point twist fields of quantum field theory (QFT), and explain their
use for the computation
of the bi-partite entanglement entropy. Branch-point twist fields were introduced in a
series of recent works in order to compute the bi-partite entanglement entropy of
1+1-dimensional QFT
\cite{entropy,other,next,nexttonext,review}. Using the ``replica trick," one writes the
entropy (\ref{defeegs}) as \beq\label{formulan1}
    S_A = -\lim_{n\to 1}\frc{d}{dn} \Tr_{{\cal A}}\left( \rho_A^n \right).
\eeq The quantity $\Tr_{{\cal
A}}\left( \rho_A^n \right)$, for integer $n$, can be interpreted as the partition function
associated to $n$ copies or replicas of the model, branched in a particular way.
In the context of 1+1-dimensional QFT, we found \cite{entropy}
that the partition function
is proportional to the
two-point function of a pair of branch-point twist fields $\mathcal{T}$,
$\tilde{\mathcal{T}}$:
\begin{equation}
   \Tr_{{\cal A}}\left( \rho_A^n
\right) \sim \langle \mathcal{T}(r) \tilde{\mathcal{T}}(0)
\rangle.
\end{equation}
These twist fields are characterized by
their exchange relations with any local field $\varphi_i$ of the $i$-th copy of the model\footnote{Here
we employ the standard notation in Minkowski space-time: $x^{\nu}$
with $\nu=0,1$, with $x^{0}$ being the time coordinate and $x^{1}$
being the position coordinate.}
\begin{eqnarray}\label{cr}
    \varphi_{i}(y)\mathcal{T}(x) &=& \Theta(x^1>y^1)\;\mathcal{T}(x) \varphi_{i+1}(y)+
        \Theta(x^1<y^1) \;\mathcal{T}(x) \varphi_{i}(y) \\
    \varphi_{i}(y)\tilde{\mathcal{T}}(x) &=& \Theta(x^1>y^1)\;\tilde{\mathcal{T}}(x) \varphi_{i-1}(y)+
        \Theta(x^1<y^1)\; \tilde{\mathcal{T}}(x) \varphi_{i}(y) \label{cr2}
\end{eqnarray}
for $i=1,\ldots, n$ and where we identify the indices $n+i \equiv
i$. Hence, branch-point twist fields are local twist fields in the
replica model associated with generating elements of its
$\mathbb{Z}_n$ symmetry. Locality plays a crucial role in
evaluating their two-point function. Of course, in order to obtain
the entanglement entropy, one performs a continuation to real $n$
and the limit $n \rightarrow 1$ in the resulting expression.

In the context of quantum spin chains, it is possible to define
cyclic permutation operators in terms of which the equivalent of
the branch-point twist fields of QFT can be constructed.
Correlation functions of such permutation operators yield the
entanglement entropy of subsets of the chain. In this context,
there is also a concept of locality associated to such permutation
operators, although we will not make explicit use of it. We will
rather show, through examples, that the use of cyclic permutation
operators provides a clear {\em combinatoric} method for
evaluating the entanglement entropy.

Second, we use this tool in order to compute the von Neumann and
R\'enyi entanglement entropies of the state obtained in the limit
$\Delta\to-1^+$ of the infinite-length anisotropic XXZ quantum
spin chain ($\Delta$ being the anisotropy parameter). The limit $\Delta\to-1^+$ and some special entanglement 
features of the associated ground state have been 
previously considered in \cite{colomo}, where the 
concurrence was computed, albeit for the case of finite 
chains. In the case of infinite chains, there are
subtleties involved in this limit. In order to properly address
them, we provide a treatment of quantum states in infinite-length
chains based on linear functionals on the space of
finitely-supported operators. Using a completely natural compact
convergence topology on linear functionals, we describe the limit
$\Delta\to-1^+$ as a proper quantum state in this framework.

We interpret our results by writing the state as a ``hybrid''
between a factorisable ground state of the XXZ chain at
$\Delta=-1$, and a state preserving the antiferromagnetic property
of XXZ ground states in the range $-1<\Delta\leq 1$ (the $z$
component of the total spin being zero). We propose the existence
of a new length scale, below which the spins behave as if they
were in a ground state of the XXZ chain at $\Delta=-1$, and above
which this antiferromagnetic property holds. As $\Delta\to-1^+$,
this scale tends to infinity, and the entanglement entropy of
finite blocks only measures the entanglement due to the
antiferromagnetic condition.

Recent results \cite{Ercolessi:2010eb} show that the entanglement
entropy between the semi-infinite halves of the infinite-length
XYZ spin chain behaves, near to certain critical lines, in ways
that are not explained by the usual CFT arguments. Although the
cases that we study do not overlap with the cases studied there,
we obtain similar conclusions. The von Neumann and R\'enyi
entanglement entropies associated to finite blocks of sites in the
$\Delta\to-1^+$ limit-state behave, as the blocks become large, in
a way that agrees with scale invariance, but not with conformal
invariance. We suggest that some results of
\cite{Ercolessi:2010eb} may be due to an interplay between the new
length scale that we propose and the usual length scale associated
to the mass gap. We hope that the relationship between our results
and those of \cite{Ercolessi:2010eb} will be made more precise in
a future publication \cite{twistagain}.

The paper is organized as follows: in Section 2 we introduce a new
type of local operators which can be defined for a replica version
of any quantum spin chain model. These are local cyclic replica
permutation operators, and we provide an explicit expression for
them in terms of elementary matrices. For completeness, we provide
proofs of their main properties, including their action on quantum
states and their exchange relations with respect to other local
operators in the chain. We explain how the bi-partite entanglement
entropy can be expressed in terms of correlation functions of such
operators. In Section 3 we apply the results of Section 2 to the
particular case of the spin-$\frac{1}{2}$ XXZ quantum spin chain
and compute the von Neumann entropy associated to one or two spins
in the infinite chain. In Section 4 we define the notion of
quantum state in infinite-length spin chains. In Section 5 we find
exact formulae for the von Neumann and R\'enyi entropies
associated to the quantum state occurring in the limit
$\Delta\to-1^+$. We study the asymptotic behavior of the entropy
for large subsystem size and find that it scales logarithmically
but not conformally. We finish this section with a discussion and
some speculation about the nature of this quantum state. We
conclude the paper in Section 6.

\section{Replica permutation operators and entanglement entropy in quantum chains}\label{sec2}

A quantum chain is a quantum system physically composed of a
number $N$ of locally-interacting sites along a one-dimensional
chain. Mathematically, two conditions are imposed. First, the
Hilbert space is ${\cal H}_N := \otimes_{i=1}^N V_i$ where $V_i
\cong \C^d$ represents the states of the individual site $i$, with
the usual inner product (we may think of this as a spin chain of
spin $(d-1)/2$)\footnote{It is not important for many of the
considerations that all sites have the same dimension; yet, for
simplicity, in most of this paper we will specialise to $d=2$.}.
Second, the Hamiltonian $H=\sum_{i=1}^N h_i$, representing the
interaction between the sites, is a sum over local energy
operators $h_i$, usually without explicit $N$ dependence. Each
$h_i$ factorises to the identity on all but a certain number
(independent of $N$) of sites. This gives rise to a notion of
``neighbourhood'': a site $j$ is in a neighbourhood of a site $i$
if $h_i$ does not factorise to the identity on site $j$. It is
this notion of neighbourhood that tells us that the chain is
actually one-dimensional, that provides a concept of distance
(e.g. the minimal length of a sequence of sequentially
intersecting neighbourhoods required to cover two given sites),
that tells us about its most natural topology (e.g. a chain with
boundaries, or a periodic chain), and that usually gives rise, in
the large-$N$ limit, to factorisation of correlation functions at
large distances. This notion of neighbourhood is also at the
source of the emergence of a local QFT in the scaling limit, when
a quantum critical point exists. The notion of neighourhood
extracted from the energy density can often be made to agree with
the usual notion of neighbourhoods on a chain (and we will assume
this throughout).

In this perspective, it is natural to look for a quantum-chain
precursors  for our branch-point twist field defined in 1+1
dimensional QFT \cite{entropy}. Recalling the exchange relations
(\ref{cr})-(\ref{cr2}), it is clear that these will be certain
permutation operators (elements of a linear representation of the
permutation group), and like in the context of QFT, the
entanglement entropy in quantum spin chains can be expressed in
terms of correlation functions of these operators.

We consider $n$ independent copies, or {\em replicas}, of a quantum chain. The Hilbert space can be described by ${\cal H}_N^{\otimes n} \cong {\cal H}_{N}^{(n)} := \otimes_{i=1}^N \otimes_{\alpha=1}^n V_{\alpha,i} = \otimes_{i=1}^N V_{i}^{(n)}$ where $V_{\alpha,i} \cong \C^d$ and the vector space at site $i$ is $V_i^{(n)} := \otimes_{\alpha=1}^n V_{\alpha,i}$. It will be convenient to denote vectors belonging to $V_i^{(n)}$ by $|s_1 s_2\ldots s_n \rangle_i,\; s_\alpha\in \{1,2,\ldots,d\}$, where the ``spin'' $s_\alpha,\;\alpha=1,\ldots,n$ belong to copy $\alpha$ of the model at that particular site $i$.

The precise permutation operators that we need, denoted $\tw_i$, are {\em local cyclic replica permutation operators}, acting non-trivially only on site $i$ in the $n$-copy spin chain, with $i=1,\ldots, N$. They act as follows:
\begin{equation}\label{ac}
\mathcal{T}_i| s_1 s_2\ldots s_n \rangle_i =| s_2 s_3\ldots s_n
s_1 \rangle_i.
\end{equation}
For $A\subset \{1,2,\ldots,N\}$ a set of sites on the chain, we will also use the notation
\beq
    \tw_A := \prod_{i\in A} \tw_i.
\eeq
Following similar ideas as in QFT, the trace of the $n$-th power of the reduced density matrix $\rho_A$ associated to a state $|\psi\ket\in{\cal H}_N$ is
\beqa
    \Tr_{\cal A}\lt(\rho_A^n\rt) &=& \Tr_{\cal B} \lt(\rho_B^n\rt) \n
        &=& \frc1{(\bra\psi|\psi\ket)^n}
            \sum_{\{|\phi_\alpha^A\ket \in{\cal A}\}\atop \{|\phi^B_\alpha\ket\in{\cal B}\}} \bra \phi_n^A\phi_n^B|\psi\ket
            \bra\psi|\phi_n^A\phi_{n-1}^B\ket \cdots \bra \phi_2^A\phi_2^B|\psi\ket
            \bra\psi|\phi_2^A\phi_1^B\ket \bra \phi_1^A\phi_1^B|\psi\ket
            \bra\psi|\phi_1^A\phi_n^B\ket \n
        &=& \frc1{\bra\Psi|\Psi\ket} \bra\Psi| \;\prod_{i\in A} \tw_i
            \sum_{\{|\phi_\alpha^A\ket \in{\cal A}\}\atop \{|\phi^B_\alpha\ket\in{\cal B}\}}
            \prod_{\alpha=1}^n |\phi_{\alpha}^A \phi_\alpha^B\ket\bra \phi_{\alpha}^A \phi_\alpha^B|\; |\Psi\ket \n
        &=&
     \frac{\langle \Psi|
    \mathcal{T}_A| \Psi \rangle }{\langle \Psi| \Psi \rangle
    }, \label{state}
\eeqa
where $|\Psi\ket = |\psi\ket^{\otimes n}$, and the sums are over orthonormal bases. Hence, the bi-partite R\'enyi and von Neumann entanglement entropies of the block $A$ can be evaluated as follows:
\begin{equation}\label{ren}
    S_{A}^{\text{R\'enyi}}(n) =\frc1{1-n} \log\lt(\frac{\langle \Psi|
    \mathcal{T}_A| \Psi \rangle }{\langle \Psi| \Psi \rangle
    }\rt).
\end{equation}
and
\begin{equation}\label{ent}
    S_{A}=-\lim_{n \rightarrow 1}\frac{d}{dn} \left(\frac{\langle \Psi|
    \mathcal{T}_A| \Psi \rangle }{\langle \Psi| \Psi \rangle
    }\right).
\end{equation}
(Naturally, any other cyclic element of the permutation group could have been used, by permutation invariance). As usual, an analytic continuation in $n$ is understood in the latter expression. We will come back below to the full relationship between these permutation operators and QFT branch-point twist fields.

\subsection{Expressions in terms of elementary matrices}

We now study the relation between these permutation operators and
elementary matrices. This will be useful because correlation
functions of the latter have tractable expressions  in the context
of integrable quantum chains. For simplicity, let us consider the
case of a spin-$\frac{1}{2}$ chains, where the spin variables
$s_1, \ldots, s_n$ above can only take two values (that is,
$d=2$). The generalisation to higher spins is straightforward and
explained below. The $2\times 2$ elementary matrices $E^{\epsilon
\epsilon'}$, for $\ep,\ep' \in \{1,2\}$, have matrix elements
given by
\begin{equation}\label{ele}
    (E^{\ep\ep'})_{kk'}=\delta_{\ep,k}\delta_{\ep',k'}.
\end{equation}
From these matrices, we form the operators
\begin{equation}\label{eleop}
    E_{\alpha,i}^{\epsilon \epsilon'},\qquad
    \alpha=1,\ldots,n \qquad \text{and} \qquad i=1,\ldots, N,
\end{equation}
which act on site $i$ and copy $\alpha$ of the quantum chain as the matrices $E^{\epsilon \epsilon'}$, and everywhere else as the identity operator.

In the simplest non-trivial situation, the two-copy model ($n=2$),
the permutation operator $\tw_i$ simply exchanges copies 1 and 2
at site $i$. It has a well-known expression in terms of elementary
matrices given by
\begin{equation}\label{2}
    \mathcal{T}_i=E_{1,i}^{11} E_{2,i}^{11}+E_{1,i}^{12}
    E_{2,i}^{21}+ E_{1,i}^{21} E_{2,i}^{12}+E_{1,i}^{22}
    E_{2,i}^{22}.
\end{equation}
It is a rather easy (although tedious) exercise to generalize ``by hand'' this
result to higher values of $n$, keeping spin-$\frac{1}{2}$. We show here just the results for
$n=3$ and $n=4$,
\begin{eqnarray}\label{3c}
     \mathcal{T}_{i}&=& E_{1,i}^{11} E_{2,i}^{11} E_{3,i}^{11} +E_{1,i}^{12}
    E_{2,i}^{21} E_{3,i}^{22} + E_{1,i}^{21} E_{2,i}^{12} E_{3,i}^{11} +E_{1,i}^{22}
    E_{2,i}^{22} E_{3,i}^{22}+\nonumber\\
     &&E_{1,i}^{11} E_{2,i}^{21}E_{3,i}^{12}+E_{1}^{12} E_{2,i}^{11}
     E_{3,i}^{21}+E_{1,i}^{22} E_{2,i}^{12}E_{3,i}^{21}+E_{1,i}^{21} E_{2,i}^{22}
     E_{3,i}^{12},
\end{eqnarray}
\begin{eqnarray}\label{4c}
     \mathcal{T}_{i}&=& E_{1,i}^{11} E_{2,i}^{11} E_{3,i}^{11}  E_{4,i}^{11} +E_{1,i}^{21} E_{2,i}^{12} E_{3,i}^{11}
     E_{4,i}^{11}+ E_{1,i}^{11} E_{2,i}^{21} E_{3,i}^{12}  E_{4,i}^{11}+E_{1,i}^{21} E_{2,i}^{22} E_{3,i}^{12}
     E_{4,i}^{11}+\nonumber\\
     &&   E_{1,i}^{11} E_{2,i}^{11} E_{3,i}^{21}  E_{4,i}^{12} + E_{1,i}^{21} E_{2,i}^{22} E_{3,i}^{22}
     E_{4,i}^{12}+ E_{1,i}^{21} E_{2,i}^{12} E_{3,i}^{21}  E_{4,i}^{12} +E_{1,i}^{11} E_{2,i}^{21} E_{3,i}^{22}
     E_{4,i}^{12}+\nonumber\\
     && E_{1,i}^{12} E_{2,i}^{11} E_{3,i}^{11}  E_{4,i}^{21} + E_{1,i}^{22} E_{2,i}^{12} E_{3,i}^{11}
     E_{4,i}^{21}+ E_{1,i}^{12} E_{2,i}^{21} E_{3,i}^{12}  E_{4,i}^{21} +E_{1,i}^{22} E_{2,i}^{22} E_{3,i}^{12}
     E_{4,i}^{21}+\nonumber\\
     && E_{1,i}^{12} E_{2,i}^{21} E_{3,i}^{22}  E_{4,i}^{22} +E_{1,i}^{22} E_{2,i}^{12} E_{3,i}^{21}
     E_{4,i}^{22}+ E_{1,i}^{12} E_{2,i}^{11} E_{3,i}^{21}  E_{4,i}^{22} +
     E_{1,i}^{22} E_{2,i}^{22} E_{3,i}^{22}
     E_{4,i}^{22}.
\end{eqnarray}
Once a few explicit cases have been worked out the general
structure of the permutation operator for generic $n$ quickly
starts to emerge. There are several features that we could have
predicted from the start. For example, all the formulae above are
sums of exactly $2^n$ terms, which is also the number of distinct
basis vectors $|s_1 s_2\ldots s_n \rangle_i$ that can be
constructed for a spin-$\frac{1}{2}$ model. The permutation
operator is a sum of $2^n$ such terms since its action (\ref{ac})
is a one-to-one map between such basis vectors. In addition,
all expressions are symmetric under the combined exchanges $E^{11}
\leftrightarrow E^{22}$ and $E^{12} \leftrightarrow E^{21}$.

One can characterize the precise $2^n$
terms that will appear in the expression for the permutation
operator through a set of four simple rules:
\begin{rules}\label{4rules}
The operator $\tw_i$ is the sum of each possible term, with coefficient 1, that is a
product of matrices
$E^{\epsilon_\alpha,\epsilon'_\alpha}_{\alpha,i}$,
$\alpha=1,\ldots,n$ respecting the following rules:
\begin{enumerate}
    \item a matrix $E_{\alpha,i}^{11}$ can only be followed in the product by
    $E^{11}_{\alpha+1,i}$ or $E^{21}_{\alpha+1,i}$,
    \item a matrix $E^{22}_{\alpha,i}$ can only be followed in the product by
    $E^{22}_{\alpha+1,i}$ or $E^{12}_{\alpha+1,i}$,
    \item a matrix $E^{12}_{\alpha,i}$ can only be followed in the  product by
    $E^{21}_{\alpha+1,i}$ or $E^{11}_{\alpha+1,i}$,
    \item  a matrix $E^{21}_{\alpha,i}$ can only be followed in the  product by
    $E^{12}_{\alpha+1,i}$ or $E^{22}_{\alpha+1,i}$
\end{enumerate}
(with $n+1\equiv 1$ and where the rules apply cyclically in the
product: the first factor follows the last factor).
\end{rules}
This restricts
greatly the type of terms that will emerge. For example
there will never be terms which involve both matrices $E^{11}$ and
$E^{22}$ and no other type, although the terms $E^{11}_{1,i}
E^{11}_{2,i}  \cdots  E^{11}_{n,i}$ and $E^{22}_{1,i} E^{22}_{2,i}
\cdots E^{22}_{n,i}$ are allowed and always appear, as we can see
for $n=2,3,4$ above. Similarly, the number of matrices $E^{12}$
and $E^{21}$ in a given product is always the same. In fact, the rules prescribe the following simple structure for the elementary matrices at any given site along the $n$ copies: strings of any number (including zero) of $E^{11}$ and $E^{22}$ are separated by $E^{21}$ (in the junctions from $E^{11}$ to $E^{22}$) and by $E^{12}$ (in the junctions from $E^{22}$ to $E^{11}$), with a condition of periodicity. This point of view will be very useful in explicit calculations below.

The rules above immediately give an expression for the permutation
operator of the form
\begin{equation}\label{pitu}
    \mathcal{T}_i=\sum_{\ep_1, \ldots \ep_{n}=1}^2 E_{1,i}^{\ep_2 \ep_1}
    E_{2,i}^{\ep_3 \ep_2 }E_{3,i}^{ \ep_4 \ep_3}\cdots  E_{n,i}^{\ep_{1}
    \ep_{n}}.
\end{equation}
This reveals the structure of the trace of some matrix. More precisely, if we introduce an auxiliary space $V_\aux \cong \C^2$ as well as the operators
\begin{equation}
  T_{\alpha,i;\aux} = \sum_{\ep,\ep'=1}^2 E^{\ep\ep'}_{\aux} E^{\ep'\ep}_{\alpha,i} = \left(
\begin{array}{cc}
  E^{11}_{\alpha,i} & E^{21}_{\alpha,i} \\
  E^{12}_{\alpha,i} & E^{22}_{\alpha,i} \\
\end{array}
\right)_\aux,
\end{equation}
which acts non-trivially both on the auxiliary space, and at site $i$, copy $\alpha$, then the permutation operator at site $i$ can be written as the
following trace
\begin{equation}\label{trace}
    \mathcal{T}_i=\text{Tr}_\aux \left(T_{1,i;\aux} T_{2,i;\aux}\cdots T_{n,i;\aux}
    \right)=\text{Tr}_\aux\left(\prod_{\alpha=1}^n T_{\alpha,i;\aux}
    \right)
\end{equation}
where the product is, by convention, ordered from left to right in order of increasing $\alpha$. It is easy to obtain (\ref{2}), (\ref{3c}) and (\ref{4c}) from
(\ref{trace}) by setting $n=2, 3$ and 4, respectively. As it should be, $\tw_i$ reduces to the identity matrix for $n=1$.

We note that the operator $T_{\alpha,i;\aux}$ is simply a
permutation operator exchanging the auxiliary space with the space
$(\alpha,i)$. The form (\ref{trace}) of the cyclic permutation
operator is something that is well-known in the context of
integrable quantum spin chains solved by means of algebraic Bethe
ansatz. In this context, the $L$-matrix is, at a certain value of
the spectral parameter, simply a permutation operator exchanging
an auxiliary space and the space associated to a given site of the
quantum chain (see, e.g., \cite{Faddeev:1996iy}). The monodromy
matrix is the trace of the product of $L$-matrices, exactly of the
form (\ref{trace}), and is identified, at this special value of
the spectral parameter, with the translation operator along the
periodic chain. In our context, however, the operator that we
obtain permutes the {\em copies} of a replica model at a given
site, rather than the sites of the chain.

In fact, this connection between our expression (\ref{trace}) and
the algebraic Bethe ansatz becomes very explicit if one employs
the solutions to the inverse scattering problem found in
\cite{JM1} to express the operators $E^{\epsilon
\epsilon'}_{\alpha,i}$ in terms of the entries of the monodromy
matrix. We suspect that this relationship could be useful in the
future for studying the properties of the replica permutation
operator within the Bethe ansatz framework.

\subsection{Exchange relations}

In the derivation above, it is clear that the Rules \ref{4rules},
the expression (\ref{pitu}), and the trace expression
(\ref{trace}) are equivalent. A careful analysis of the
permutation action (\ref{ac}) would also give rise to the four
rules stated. Here, for completeness, we will show explicitly that
equation (\ref{ac}) is a consequence of (\ref{trace}). In order to
do so, we will show that the main exchange property of replica
permutation operators (the property that constitutes the starting
point for the definition of twist fields in QFT) is satisfied by
the operator (\ref{trace}).

Given a 2 by 2 matrix $\mathcal{O}$, let $\mathcal{O}_{\alpha,i}$ be the operator on the quantum chain that acts as $\mathcal{O}$ on site $i$, copy $\alpha$, and as the identity operator everywhere else. Then, we have the following:
\begin{lemma}
The operators $\tw_i$ defined by (\ref{trace}) satisfy the relations
\begin{equation}
\mathcal{T}_i \mathcal{O}_{\alpha,i} =
\mathcal{O}_{\alpha-1,i}\mathcal{T}_i \label{exchange}
\end{equation}
(with $\mathcal{O}_{0,i} \equiv \mathcal{O}_{n,i}$) for all $i=1,\ldots,N$, all $\alpha=1,\ldots,n$, and all $\mathcal{O}$.
\end{lemma}
\proof This relation follows from the trace expression (\ref{trace}) along with two identities. Denoting by $\mathcal{O}_{\aux}$ the operator acting non-trivially as $\mathcal{O}$ on the auxiliary space and as the identity operator everywhere else, we only have to show that
\beq
    {T}_{\alpha,i;\text{aux}}\mathcal{O}_{\alpha,i} = \mathcal{O}_{\text{aux}} {T}_{\alpha,i;\text{aux}},\quad
    {T}_{\alpha,i;\text{aux}}\mathcal{O}_{\aux} = \mathcal{O}_{\alpha,i} {T}_{\alpha,i;\text{aux}}.
\label{exchange2}
\eeq
The first identity can be derived as follows. With
\begin{equation}
 \mathcal{O} = \left(
\begin{array}{cc}
  o_{11} & o_{12} \\
 o_{21} & o_{22}\\
\end{array}
\right),
\end{equation}
we have
\begin{eqnarray}
{T}_{\alpha,i;\text{aux}}\mathcal{O}_{\alpha,i}&=&
\left(
\begin{array}{cc}
  E^{11}_{\alpha,i} & E^{21}_{\alpha,i} \\
  E^{12}_{\alpha,i} & E^{22}_{\alpha,i} \\
\end{array}
\right)_{\aux} \left(
\begin{array}{cc}
 \mathcal{O}_{\alpha,i} & 0 \\
 0 & \mathcal{O}_{\alpha,i} \\
\end{array}
\right)_\text{aux}= \left(
\begin{array}{cc}
  E^{11}_{\alpha,i} \mathcal{O}_{\alpha,i} & E^{21}_{\alpha,i} \mathcal{O}_{\alpha,i} \\
  E^{12}_{\alpha,i} \mathcal{O}_{\alpha,i} & E^{22}_{\alpha,i}  \mathcal{O}_{\alpha,i}\\
\end{array}
\right)_\text{aux}\nonumber\\ &=&  \left(
\begin{array}{cc}
 o_{11} E^{11}_{\alpha,i}+o_{12} E^{21}_{\alpha,i} & o_{11}E^{21}_{\alpha,i}+o_{12}E^{22}_{\alpha,i} \\
 o_{21} E^{11}_{\alpha,i}+o_{22} E^{12}_{\alpha,i}  & o_{21} E^{21}_{\alpha,i}+o_{22} E^{22}_{\alpha,i} \\
\end{array}
\right)_\text{aux},
\end{eqnarray}
which is equal to
\begin{eqnarray}
 \mathcal{O}_{\aux} {T}_{\alpha,i;\text{aux}}&=&  \left(
\begin{array}{cc}
  o_{11} & o_{12} \\
 o_{21} & o_{22}\\
\end{array}
\right)_\text{aux}  \left(
\begin{array}{cc}
  E^{11}_{\alpha,i} & E^{21}_{\alpha,i} \\
  E^{12}_{\alpha,i} & E^{22}_{\alpha,i} \\
\end{array}
\right)_\text{aux}  \nonumber\\
&=&  \left(
\begin{array}{cc}
 o_{11} E^{11}_{\alpha,i}+o_{12} E^{21}_{\alpha,i} & o_{11}E^{21}_{\alpha,i}+o_{12}E^{22}_{\alpha,i} \\
 o_{21} E^{11}_{\alpha,i}+o_{22} E^{12}_{\alpha,i}  & o_{21} E^{21}_{\alpha,i}+o_{22} E^{22}_{\alpha,i} \\
\end{array}
\right)_\text{aux}.
\end{eqnarray}
The second relation of (\ref{exchange2}) follows in a similar way.
\eproof

The defining property (\ref{ac}) is essentially a consequence of
(\ref{exchange}).
\begin{lemma}
The operators $\tw_i$ defined (\ref{trace}) satisfy the relations (\ref{ac}).
\end{lemma}
\proof We start by writing the quantum state of the
spin-$\frac{1}{2}$, $n$-copy quantum spin chain at site $i$ in
terms of elementary matrices $E_{\alpha,i}^{\epsilon_1
\epsilon_2}$. Let $|0\rangle_i$ be a reference state at site $i$
for which all spins of all $n$-copies are up
($s_1=\cdots=s_n=\uparrow$). Any other of the $2^n$ possible spin
configurations can be generated by acting on this reference state
with a combination of elementary matrices. For example, if we want
to lower the spin of copy 1, we just have to compute
$E^{21}_{1,i}|0\rangle_i$, whereas acting with $E^{11}_{1,i}$ will
leave the state unchanged and the action of $E^{22}_{1,i}$ and
$E^{12}_{1,i}$ gives zero. Therefore, a generic state is written as
\begin{equation}
  |s_1 s_2\ldots s_n\rangle_i=\left(\prod_{\alpha=1}^n E^{j_{s_\alpha} 1}_{\alpha,i}\right)
  |0\rangle_i, \quad j_{s_\alpha}=1 \quad \text{for} \quad
  s_\alpha=\uparrow\quad \text{and}
  \quad j_{s_\alpha}=2 \quad \text{for} \quad s_\alpha=\downarrow,
\end{equation}
and from the expression (\ref{pitu}), we have $\tw_i|0\ket = |0\ket$ since the only non-zero summand is the one where $\ep_\alpha=1$ for all $\alpha$. We now  evaluate the action of the permutation operator on this
state by successively employing (\ref{exchange}),
\begin{eqnarray}
 \mathcal{T}_i |s_1,s_2,\ldots,s_n\rangle_i&=&\mathcal{T}_i
 \left(\prod_{\alpha=1}^n E^{j_{s_\alpha} 1}_{\alpha,i}\right)
  |0\rangle_i\nonumber\\&=&\left(\prod_{\alpha=1}^n E^{j_{s_\alpha} 1}_{\alpha-1,i}\right)
  \mathcal{T}_i
  |0\rangle_i =\left(\prod_{\alpha=1}^n E^{j_{s_\alpha} 1}_{\alpha-1,i}\right)
  |0\rangle_i=|s_2,\ldots,s_n,s_1\rangle_i.
\end{eqnarray}
This establishes (\ref{ac}).
\eproof

\subsection{The conjugate permutation operator $\tilde{\mathcal{T}}$}

Other permutation operators, $\tilde{\mathcal{T}}_i$ at each site
$i$, take us from copy
  $\alpha$ to copy $\alpha-1$. We
     can simply obtain $\tilde{\mathcal{T}}_i$ by reversing the order of the replicas, that is
\begin{equation}
\label{trace2}
   \tilde{\mathcal{T}}_i=\text{Tr}_{\text{aux}}\left(T_{n,i;\text{aux}} T_{n-1,i;\text{aux}}
   \cdots T_{1,i;\text{aux}}
    \right),
\end{equation}
or, alternatively, taking the trace,
\begin{equation}
\label{pitu3}
   \tilde{\mathcal{T}}_i=\sum_{\epsilon_1, \ldots \epsilon_{n}=1}^2 E_{n,i}^{\epsilon_2 \epsilon_1}
    E_{n-1,i}^{\epsilon_3 \epsilon_2 }E_{n-2,i}^{ \epsilon_4 \epsilon_3}\cdots  E_{1,i}^{\epsilon_{1}
    \epsilon_{n}},
\end{equation}
One may check this explicitly by showing that $
\tilde{\mathcal{T}}_i   {\mathcal{T}}_i ={\bf 1}$. This can be
done by using the following property of the elementary matrices:
\begin{equation}
E^{\epsilon_1, \epsilon_2}_{\alpha,i} E^{\epsilon_3
\epsilon_4}_{\alpha,i} = \delta_{\epsilon_2, \epsilon_3}
E^{\epsilon_1 \epsilon_4}_{\alpha,i}. \label{pitu2}
\end{equation}
Employing the representations (\ref{pitu}) and (\ref{pitu3}) for
the permutation operators (including some extra matrices in the
products for clarity) we have
\begin{eqnarray}
   \tilde{\mathcal{T}}_i   {\mathcal{T}}_i  &=&  \sum_{\epsilon_1, \ldots \epsilon_{n}=1}^2 E_{n,i}^{\epsilon_2 \epsilon_1}
    E_{n-1,i}^{\epsilon_3 \epsilon_2}E_{n-2,i}^{ \epsilon_4 \epsilon_3}\cdots E_{3,i}^{\epsilon_{n-1} \epsilon_{n-2}} E_{2,i}^{\epsilon_{n}\epsilon_{n-1}}
     E_{1,i}^{\epsilon_{1}
    \epsilon_{n}}\nonumber \\
    && \times \sum_{\epsilon'_1, \ldots \epsilon'_{n}=1}^2 E_{1,i}^{\epsilon'_2 \epsilon'_1}
    E_{2,i}^{\epsilon'_3 \epsilon'_2 }E_{3,i}^{ \epsilon'_4 \epsilon'_3}\cdots E_{n-2,i}^{\epsilon'_{n-1}
    \epsilon'_{n-2}} E_{n-1,i}^{\epsilon'_{n}
    \epsilon'_{n-1}} E_{n,i}^{\epsilon'_{1}
    \epsilon'_{n}}\nonumber\\
    &=& \sum_{\epsilon_1, \ldots \epsilon_{n}=1}^2 \sum_{\epsilon'_1, \ldots \epsilon'_{n}=1}^2
    \delta_{\epsilon_1,\epsilon'_1}  E_{n,i}^{\epsilon_2 \epsilon'_n} \delta_{\epsilon_2 \epsilon'_n} E_{n-1,i}^{\epsilon_3 \epsilon'_{n-1}}
    \ldots \delta_{\epsilon_{n-1} \epsilon'_3} E_{2,i}^{\epsilon_n \epsilon'_2} \delta_{\epsilon_n \epsilon'_2}E_{1,i}^{\epsilon_1 \epsilon'_1}\nonumber\\
    &=& \sum_{\epsilon_1, \ldots \epsilon_{n}=1}^2 E_{n,i}^{\epsilon_2 \epsilon_2}
    E_{n-1,i}^{\epsilon_3 \epsilon_3} E_{n-2,i}^{\epsilon_4 \epsilon_4} \ldots E_{3,i}^{\epsilon_{n-1} \epsilon_{n-1}}
    E_{2,i}^{\epsilon_n \epsilon_n} E_{1,i}^{\epsilon_1 \epsilon_1}={\bf 1}.
\end{eqnarray}
The last equality follows from the fact that all matrices
$E_{\alpha,i}^{\epsilon \epsilon}$ are diagonal, so that each
product above is a diagonal matrix with a single non-vanishing
entry 1. The sum over all such matrices (there are exactly $2^n$
of them) is the identity.

\subsection{Replica permutation operators for higher spin chains}

We can now generalize the results above to more general spin
chains, with arbitrary spin representations living at each site of
the chain. Indeed the exchange relations (\ref{exchange}) which we
have just proven can be employed as a starting point for finding a
realization of the permutation operator which is valid for higher
spin chains. It can be shown that, assuming a dimension-$d$
representation lives at site $i$ of the quantum spin chain, the
permutation operator $\mathcal{T}_i$ can be expressed exactly as
in (\ref{trace}) with  $T_{\alpha,i;\text{aux}}$ given by
\begin{equation}
  T_{\alpha,i;\text{aux}}=  \left(
\begin{array}{cccc}
  E^{11}_{\alpha,i} & E^{21}_{\alpha,i}&\cdots & E^{d1}_{\alpha,i} \\
  E^{12}_{\alpha,i} & E^{22}_{\alpha,i} &\cdots & E^{d2}_{\alpha,i} \\
 \vdots  & \vdots & \ddots &\vdots \\
  E^{1d}_{\alpha,i} & E^{2d}_{\alpha,i} &\cdots & E^{dd}_{\alpha,i} \\
\end{array}
\right)_\text{aux}. \label{gent}
\end{equation}
It is not too difficult to show that all properties established in
the previous section for spin-$\frac{1}{2}$ also hold for the
general case. We will not dwell on such proofs here as in the
remainder of this paper we will concentrate on the
spin-$\frac{1}{2}$ case.

\subsection{Local-unitary operators and twist fields}

Since the operator $\mathcal{T}_i$ is real, we immediately find,
from (\ref{trace}) and (\ref{trace2}), that
\[
    \tilde{\mathcal{T}}_i= \mathcal{T}_i^\dagger.
\]
Hence, $\tw_i$ is a unitary operator. It is of course the local
element of the unitary operator associated with the cyclic replica
permutation symmetry of the $n$-copy Hamiltonian \beq\label{Hn}
    H^{(n)} = \sum_{i=1}^N h_i^{(n)} = \sum_{i=1}^N \sum_{\alpha=1}^n h_{\alpha,i}
\eeq
(where $h_{\alpha,i}$, for $\alpha=1,\ldots,n$ and $i=1,\ldots,N$, acts as $h_i$ on $\otimes_{j=1}^N V_{\alpha,j}$, and as the identity on $\otimes_{j=1}^N V_{\alpha',j}\ \forall\ \alpha'\neq\alpha$). More precisely, the unitary operator that is the product of $\tw_i$ over all sites $i$, i.e. $\tw_{\{1,\ldots,N\}} = \prod_{i=1}^N \mathcal{T}_i$, implements the cyclic replica permutation of the chain, and commutes with the $n$-copy Hamiltonian. A unitary operator that can be written as a product of single-site unitary operators is sometimes called a {\em local-unitary} operator.

In fact, the local-unitary operator $\tw_{\{1,\ldots,N\}}$ commutes also with the energy density,
\begin{equation}
\lt[\tw_{\{1,\ldots,N\}},h_{i}^{(n)}\rt] = 0.
\end{equation}
In general, it is expected that any unitary operator associated to an internal symmetry, i.e. that is local-unitary, commutes not only with the Hamiltonian, but also with the energy density. To such an internal symmetry, one may then associate {\em twist fields}; e.g. in the case of $\tw_i$,
\[
    {\tt T}_i = \tw_{\{i,i+1,\ldots,N\}} = \prod_{j=i}^N \mathcal{T}_i.
\]
A standard example is the twist field involved in the construction of fermion operators on the Ising spin chain. Quantum chain twist fields give rise to the usual QFT twist fields in the scaling limit. Twist fields have the property of being local ``in the bulk'': their commutators with the energy density vanish for large enough separations (and away from the end of the chain $N$), by virtue of their association with an internal symmetry transformation. In the case of the replica permutation symmetry:
\beq\label{loc}
    [{\tt T}_i,h_{j}^{(n)}] = 0 \quad \mbox{for $j$ far enough from $i$ and from $N$.}
\eeq This has important implications in QFT, and is likely to have
important implications as well in the context of  integrable
quantum spin chains. We will not investigate further here such
implications.

Consider the permutation operators $\tw_i$ and their twist fields ${\tt T}_i$. In the scaling limit, this of course gives rise to branch-point twist fields, associated with branch-point singularities on the surface where the QFT is defined. Like in QFT, the entanglement entropy (and likewise the R\'enyi entropy) of a region $A$ of the quantum chain can be expressed via a correlation function of as many pairs of twist fields ${\tt T}_i$, ${\tt T}_i^\dag$ as there are connected sub-regions. For instance, for a single connected region $A$ starting at site $i$ and ending at site $j$, we have
\[
    S_A^{\text{R\'enyi}} = \frc1{1-n}\log \lt(\frc{\bra \Psi|{\tt T}_i {\tt T}_j^\dag|\Psi\ket}{\bra \Psi|\Psi\ket}\rt).
\]

\subsection{Local-unitary invariance}

One of the properties of the entanglement and R\'enyi entropies is
that they are invariant under local-unitary transformations
(transformation of the state by a local-unitary operator). The
precursor of this invariance is of course that the replica
permutation operators $\tw_A$ are invariant under tensor-powers of
local-unitary transformations\footnote{The set of (products of)
replica permutation operators associated with all permutation
elements spans the space of local-unitarily invariant operators
(but these operators are not all linearly independent).},
$U^{\otimes n} \tw_A (U^\dag)^{\otimes n} = \tw_A$ for $U$ a
local-unitary operator on ${\cal H}$. In general, we expect that
if two quantum states have the same R\'enyi entropies, then they
are related by a local-unitary transformation.

\section{The XXZ Heisenberg spin-$\frac{1}{2}$ chain: main features}

In order to illustrate the technique proposed, we now choose the
XXZ Heisenberg spin-$\frac{1}{2}$ chain as a benchmark. It is one
of the most studied quantum chains and includes other interesting
theories as special cases. The XXZ Heisenberg spin-$\frac{1}{2}$
finite chain of length $N$ is characterized by the Hamiltonian
\begin{equation}\label{h}
    H_\Delta=\sum_{j=1}^N \left(\sigma^x_j \sigma^x_{j+1}+\sigma^y_j \sigma^y_{j+1} + \Delta(\sigma^z_j \sigma^z_{j+1}-1)
    \right),
\end{equation}
where $\sigma_j^{x,y,z}$ are the Pauli matrices associated to site
$j$ of the chain and acting on the two-dimensional space $V_j
\cong \mathbb{C}^2$, a spin-$1/2$ module. We consider periodic
boundary conditions
$\sigma_{N+1}^{x,y,z}\equiv\sigma_{1}^{x,y,z}$, and for simplicity
we restrict to $N$ being even. The parameter $\Delta$ is known as
the anisotropy parameter and, depending on its value, the physical
properties of the model can change dramatically. In this paper, we
will be looking at the region $-1<\Delta\leq 1$. In this region,
the model is antiferromagnetic and preserves the $z$-component of
the total spin: its ground state has total $z$-spin equal to zero.
As $N \rightarrow \infty$, the model is massless: the gap between
the unique ground state and the first excited state tends to 0.
For $\Delta=1$ the model becomes what is known as the XXX model or
Heisenberg spin chain, whereas for $\Delta=0$ it reduces to the XX
model, which possesses a free fermion description. For
$\Delta=-1$, the ground state of the model is very different. It
is unitarily related to the ground state of the ferromagnetic
Heisenberg spin chain, and is not unique. In all cases, the ground
state is invariant under translations (cyclic permutations of the
sites along the chain). The anisotropy parameter is commonly
expressed in terms of another parameter $\eta$ as
\begin{equation}\label{eta}
    \Delta=\cosh(\eta),
\end{equation}
with $0\leq \eta<\pi$.

The characterization of the physical states, including the ground
state, of this and many other integrable quantum spin models can
be carried out very effectively by means of the algebraic Bethe
ansatz approach \cite{Faddeev:1979gh,Faddeev:1996iy}. The details
of the approach can be found in many places (a particularly good
review can be found in \cite{Faddeev:1996iy}) and we will not
review those here. The entanglement entropy is a
function of the correlation functions of the model. The evaluation of correlation functions
has been performed in the literature by various methods: using the algebraic Bethe ansatz
\cite{FST,Faddeev:1996iy,korepinbook} with the solution to the inverse scattering problem
\cite{JM1,JM0,KMT}, and through a computation of form factors \cite{JMi2,JMi,JMi3} in quantum
spin chains of infinite length via $q$-deformed vertex operators.

We will denote by $|\psi\ket_{\Delta}$ the ground state of the Hamiltonian (\ref{h}). We will also use the notation $E_i^{\ep\ep'}$ for the operator on ${\cal H}$ acting as the elementary matrix (\ref{ele}) at site $i$ and as the identity elsewhere. Instead of considering correlation functions of Pauli matrices, we will consider {\em fundamental blocks}, correlation functions of (i.e. quantum averages of products of) elementary matrices in the ground state $|\psi\rangle_\Delta$. We will use the following notation (with implicit dependence on $\Delta$):
\begin{equation}\label{cf}
    \langle \prod_{i\in A} E^{\epsilon_i \epsilon'_i}_{i}
    \rangle := \frac{{}_\Delta\langle \psi| \prod\limits_{i\in A}
E^{\epsilon_{i},\epsilon'_{i}}_{i} | \psi \rangle_\Delta
}{{}_\Delta\langle \psi| \psi
    \rangle_\Delta},
\end{equation}
where $A\subset \{1,2,\ldots,N\}$ is an index set.

We will denote by $|\Psi \ket$ (with implicit dependence on
$\Delta$) the vector in the replica model ${\cal H}^{(n)}$
corresponding to the tensor product $|\psi\ket^{\otimes n}_\Delta$
of the ground state. This is of course the ground state of the
$n$-copy Hamiltonian $H^{(n)}$. By factorisation, we have, for
instance,
\begin{equation}
   \frac{\langle \Psi| E^{\epsilon_1 \epsilon'_1}_{1,i} E^{\epsilon_2 \epsilon'_2}_{2,i}\ldots E^{\epsilon_n \epsilon'_n}_{n,i} | \Psi\rangle }{\langle \Psi| \Psi \rangle
    }=\prod_{\alpha=1}^n \langle E^{\epsilon_\alpha \epsilon'_\alpha}_{i}
    \rangle.\label{factor}
\end{equation}
We note that it is a simple matter to verify the general locality property (\ref{loc}) of the twist fields ${\tt T}_i$ in the present case. With $h_j = \sigma^x_j \sigma^x_{j+1}+\sigma^y_j \sigma^y_{j+1} + \Delta(\sigma^z_j \sigma^z_{j+1}-1)$ and $h_j^{(n)}$ defined as in (\ref{Hn}), we find
\[
    [{\tt T}_i,h_j^{(n)}] = \delta_{i-1,j} {\tt T}_i \sum_{\alpha=1}^n \lt(
    \sigma^x_{\alpha,i-1} (\sigma^x_{\alpha,i} - \sigma^x_{\alpha+1,i})
    + \sigma^y_{\alpha,i-1} (\sigma^y_{\alpha,i} - \sigma^y_{\alpha+1,i})
    + \Delta \sigma^z_{\alpha,i-1} (\sigma^z_{\alpha,i} - \sigma^z_{\alpha+1,i}) \rt)
\]
for $j\neq N$.

\subsection{Entanglement entropy of one and two sites}

We now perform simple computations of the entanglement entropy between one
or two sites and the rest of the chain using the permutation operator technique.
This illustrates how the technique directly gives the combinatorics for evaluating
the trace of the reduced density matrix in terms of correlation functions (avoiding
the explicit diagonalisation of the matrix).

The one-site computation can be thought of as a consistency check
of our formalism, since the outcome is known a priori. This is
because the ground state of the spin-$\frac{1}{2}$ XXZ chain is
characterized by equal numbers of spins up and down. Hence if we
pick one spin in the chain, its projection in the $z$-direction is
entirely determined by the values of that projection for the
remaining spins in the chain. In other words, the entanglement
between one spin and the rest is maximal and it is well known that
this maximum value is exactly $\log(2)$.

The entropy that we want to compute is
\begin{equation}\label{ent1}
    S_{\{i\}}=-\lim_{n \rightarrow 1}\frac{d}{dn} \left(\frac{\langle \Psi| \mathcal{T}_i| \Psi \rangle }{\langle \Psi| \Psi\rangle
    }\right).
\end{equation}
In order to evaluate this, we note that
\begin{equation}
\langle E^{12}_i\rangle= \langle E^{21}_i \rangle=0, \label{zero}
\end{equation}
and
\begin{equation}
 \langle
E^{11}_i\rangle =\langle E^{22}_i \rangle=\frac{1}{2}. \label{12}
\end{equation}
The first equality (\ref{zero}) is due to the fact that the
operators $E^{12}_i$ and $E^{21}_i$ have the effect of reversing
the spin at site $i$, turning the quantum average above into
the scalar product of the ground state with another orthogonal
state. The second equality (\ref{12}) is due to the fact that the
quantum averages involved represent the probability of
finding the spin at site $i$ up, and that of finding it down. These probabilities are
exactly $\frac{1}{2}$ due to the nature of the ground state.

Factorisation (\ref{factor}) and the expression (\ref{pitu}) mean that we have to evaluate a sum of products of one-point functions.
Because of (\ref{zero}),
the only non-vanishing contributions to the
quantum average of $\mathcal{T}_i$ will come from the two
terms that involve only matrices $E^{11}$ or $E^{22}$.
That is:
\begin{equation}\label{spin1}
\frac{\langle \Psi| \mathcal{T}_i| \Psi \rangle }{\langle
\Psi| \Psi \rangle}=\langle E^{11}_i\rangle^n +\langle
E^{22}_i\rangle^n= 2^{1-n}.
\end{equation}
The entanglement entropy is therefore, as expected
\begin{equation}\label{s1}
    S_{\{i\}}=-\lim_{n\rightarrow 1}\frac{d}{dn}2^{1-n}=\lim_{n\rightarrow
    1} 2^{1-n} \log(2)=\log(2).
\end{equation}

We now consider the computation of the entanglement
entropy of spins sitting at sites 1 and $m+1$ with respect to the
rest of the chain:
\begin{equation}
    S_{\{1,m+1\}}=-\lim_{n \rightarrow 1}\frac{d}{dn} \left(\frac{\langle \Psi|
     \mathcal{T}_1 \mathcal{T}_{m+1}| \Psi \rangle }{\langle \Psi| \Psi \rangle
    }\right)
\end{equation}
(this is the same as $S_{\{i,m+i\}}$ thanks to translation invariance).

Given the factorization property (\ref{factor}), we know that
$S_{\{1,m+1\}}$ will be given in terms of two-point functions involving
pairs of operators
$E_1^{\epsilon_1\epsilon'_1}E_{m+1}^{\epsilon_2,\epsilon'_2}$.
However, as before, many of these correlation functions are
vanishing. More precisely, they do so whenever an operator
$E^{12}$ or $E^{21}$ appears in combination with $E^{11}$ or
$E^{22}$. There are only six non-vanishing two-point
functions:
\begin{equation}
\langle E^{11}_1 E^{11}_{m+1}\rangle=\langle E^{22}_1
E^{22}_{m+1}\rangle, \qquad \langle E^{11}_1 E^{22}_{m+1}\rangle=
\langle E^{22}_1 E^{11}_{m+1}\rangle,\label{eq}
\end{equation}
and
\begin{equation}
\langle E^{12}_1 E^{21}_{m+1}\rangle=\langle E^{21}_1
E^{12}_{m+1}\rangle, \label{eq2}
\end{equation}
where the equalities are again due to spin reversal symmetry. This
means that although the operator $\mathcal{T}_1 \mathcal{T}_{m+1}$
is a sum of $4^n$ terms, each of which involving $2n$ matrices
$E^{ij}$, inside the correlation function most of these terms will
give a vanishing contribution.

We may evaluate the correlation function by using Rules
\ref{4rules}, recalling the structure mentioned there. Here, we
may consider a square lattice of cylindrical topology, with 2
columns, $n$ rows, and periodicity along the columns. Each cell is
occupied by an elementary matrix: this represents a term in the
product $\mathcal{T}_1 \mathcal{T}_{m+1}$. Rules \ref{4rules}
imply that in order to determine a column, we only need to provide
the positions of matrices $E^{12}$ and $E^{21}$, with the
condition of alternation, unless there are none. In the latter
case, there are only two possibilities: whether the column is
filled with $E^{11}$, or it is filled with $E^{22}$. On the other
hand, the constraints about non-vanishing correlation functions
imply that on a row, if a $E^{12}$ is present then a $E^{21}$ must
also be present. This implies that in order to determine a term,
we only need to determine the positions of matrices $E^{12}$ and
$E^{21}$ along one column only -- say the first column -- unless
there are none.

Suppose that along the first column there are $s$ matrices $E^{22}$, and $q$ pairs of matrices $E^{12}$ and $E^{21}$. Then there are $n-s-2q$ matrices $E^{11}$. In the case where $q=0$, by the discussion above not all values of $s$ are available, and there are exactly four terms:
\[
    \bra E^{11}_1 E^{11}_{m+1}\ket^n + \bra E^{22}_1 E^{22}_{m+1}\ket^n +
    \bra E^{11}_1 E^{22}_{m+1}\ket^n + \bra E^{22}_1 E^{11}_{m+1}\ket^n.
\]
For $q>0$, all values $q$ from 1 to $[n/2]$ (where $[\cdot]$ means integer part) occur, and all values of $s$ from 0 to $n-2q$ occur. By factorisation, every term characterised by such $s$ and $q$ gives rise to the same product of correlation functions:
\[
    \langle E^{11}_1
    E^{22}_{m+1}\rangle^{n-s-2q}\langle E^{22}_1
    E^{11}_{m+1}\rangle^{s} \langle E^{12}_1
    E^{21}_{m+1}\rangle^{q}\langle E^{21}_1
    E^{12}_{m+1}\rangle^{q}.
\]
In order to count the number of such terms, consider two possibilities: whether the jump from $n$ to 1 along the column breaks a string of (possibly a vanishing number of) matrices $E^{22}$, or it breaks a string of matrices $E^{11}$. In the first case, the number of terms is $p_{q+1}(s) p_q(n-s-2q)$, and in the second, it is $p_{q+1}(n-s-2q) p_q(s)$, where $p_q(s)$ represents the number of partitions of $s$ into a sum of $q$ parts, with the number 0 is also included as a possible part. Hence, the number of such terms is
\begin{equation}\label{coef}
    C_n(s,q)=p_{q+1}(s) p_q(n-s-2q)+p_{q+1}(n-s-2q) p_q(s).
\end{equation}
A formula for $p_q(s)$ can be easily obtained by noticing that the
function $(1-x)^{-n}$ is precisely the generating function of such
coefficients, namely
\begin{equation}
\frac{1}{(1-x)^q}=\sum_{s=0}^{\infty} p_q(s) x^s =
\sum_{s=0}^{\infty}  \left(\begin{array}{c}
  q+s-1 \\
  q-1\\
\end{array}\right) x^s,
\end{equation}
where $ \left(\begin{array}{c}
  a \\
  b\\
\end{array}\right)=\frac{a!}{b!(a-b)!}$ is the binomial
coefficient. We can easily show that the coefficient (\ref{coef})
can be written as
\begin{equation}
    C_n(s,q)=\frac{n}{n-s-q} \left(\begin{array}{c}
  q+s-1 \\
  q-1\\
\end{array}\right)  \left(\begin{array}{c}
  n-s-q \\
  q\\
\end{array}\right).
\end{equation}
Employing this result and from the discussion above, we find that
the correlation function that we want to compute is
\begin{eqnarray}
   \frac{\langle \Psi|
     \mathcal{T}_1 \mathcal{T}_{m+1}| \Psi \rangle }{\langle \Psi| \Psi \rangle
    }&=&\sum_{q=1}^{[\frac{n}{2}]}\sum_{s=0}^{n-2q} C_n(q,s) \langle E^{11}_1
    E^{22}_{m+1}\rangle^{n-s-2q}\langle E^{22}_1
    E^{11}_{m+1}\rangle^{s} \langle E^{12}_1
    E^{21}_{m+1}\rangle^{q}\langle E^{21}_1
    E^{12}_{m+1}\rangle^{q}\nonumber\\
    && + \langle E^{11}_1
    E^{11}_{m+1}\rangle^{n} +\langle E^{22}_1
    E^{22}_{m+1}\rangle^{n} + \langle E^{11}_1
    E^{22}_{m+1}\rangle^{n} + \langle E^{22}_1
    E^{11}_{m+1}\rangle^{n}.\label{spin2}
\end{eqnarray}

The
sum above is relatively complicated to compute. However, it
simplifies greatly if we employ the equalities
(\ref{eq})-(\ref{eq2}). We then obtain
\begin{eqnarray}
   \frac{\langle \Psi|
     \mathcal{T}_1 \mathcal{T}_{m+1}| \Psi \rangle }{\langle \Psi| \Psi \rangle
    }&=&\sum_{q=1}^{[\frac{n}{2}]}\langle E^{11}_1
    E^{22}_{m+1}\rangle^{n-2q}\langle E^{12}_1
    E^{21}_{m+1}\rangle^{2q}\sum_{s=0}^{n-2q} C_n(q,s) \nonumber\\
    && + 2\langle E^{11}_1
    E^{11}_{m+1}\rangle^{n}  +2 \langle E^{11}_1
    E^{22}_{m+1}\rangle^{n}.
\end{eqnarray}
It is now possible to carry out the sum. The sum in $s$ yields
simply
\begin{equation}
    \sum_{s=0}^{n-2q} C_n(q,s) =2\left(\begin{array}{c}
  n \\
  2q\\
\end{array}\right).
\end{equation}
Finally, the sum in $q$ gives
\begin{eqnarray}
   && 2 \sum_{q=1}^{[\frac{n}{2}]}\left(\begin{array}{c}
  n \\
  2q\\
\end{array}\right)\langle E^{11}_1
    E^{22}_{m+1}\rangle^{n-2q}\langle E^{12}_1
    E^{21}_{m+1}\rangle^{2q}\nonumber\\
    && =-2\langle E^{11}_1
    E^{22}_{m+1}\rangle^n+\left(\langle E^{11}_1
    E^{22}_{m+1}\rangle+{\langle E^{12}_1
    E^{21}_{m+1}\rangle}\right)^n +  \left(\langle E^{11}_1
    E^{22}_{m+1}\rangle-{\langle E^{12}_1
    E^{21}_{m+1}\rangle}\right)^n.
\end{eqnarray}
Therefore, the final expression for the two-point function of
permutation operators is
\begin{equation}\label{res2spin}
    \frac{\langle \Psi|
     \mathcal{T}_1 \mathcal{T}_{m+1}| \Psi \rangle }{\langle \Psi| \Psi \rangle
    }=\left(\langle E^{11}_1
    E^{22}_{m+1}\rangle+{\langle E^{12}_1
    E^{21}_{m+1}\rangle}\right)^n +  \left(\langle E^{11}_1
    E^{22}_{m+1}\rangle-{\langle E^{12}_1
    E^{21}_{m+1}\rangle}\right)^n + 2\langle E^{11}_1
    E^{11}_{m+1}\rangle^n,
\end{equation}
and the entropy becomes
\begin{eqnarray}
   S_{\{1,m+1\}}&=&-(\langle E^{11}_1
    E^{22}_{m+1}\rangle+{\langle E^{12}_1
    E^{21}_{m+1}\rangle})\log\left[\langle E^{11}_1
    E^{22}_{m+1}\rangle+{\langle E^{12}_1
    E^{21}_{m+1}\rangle}\right]\nonumber \\
    && -(\langle E^{11}_1
    E^{22}_{m+1}\rangle-{\langle E^{12}_1
    E^{21}_{m+1}\rangle})\log\left[\langle E^{11}_1
    E^{22}_{m+1}\rangle-{\langle E^{12}_1
    E^{21}_{m+1}\rangle}\right]\nonumber\\
    && -2\langle E^{11}_1
    E^{11}_{m+1}\rangle \log\langle E^{11}_1
    E^{11}_{m+1}\rangle.
\end{eqnarray}
Note that in (\ref{res2spin}) we can immediately read-off the eigenvalues of the reduced density matrix (the three quantities that are taken to the power $n$) and their degeneracies (the integer coefficients of these powers).

It is simple to rewrite this expression in terms of Pauli
matrices. Employing the short-hand notation $\langle \sigma^{z}_1
    \sigma^{z}_{m+1}\rangle=z(m)$  and $\langle \sigma^{+}_1
    \sigma^{-}_{m+1}\rangle=s(m)$, we have
\begin{eqnarray}\label{re}
    S_{\{1,m+1\}}
     &=& -\frac{(1-z(m)+4 s(m))}{4}\log\left[\frac{(1-z(m)+4 s(m))}{4}\right]\nonumber\\
     && -\frac{(1-z(m)-4 s(m))}{4}
     \log\left[\frac{(1-z(m)-4 s(m))}{4}\right]\nonumber\\
     && -\frac{(1+z(m))}{2}\log\left[\frac{1+z(m)}{4}\right].
\end{eqnarray}

From here let us specialise to the case of infinite length, $N\to\infty$. This makes sense because correlation functions of local operators have finite limits. Then, one feature of the expression above is that for $m\rightarrow \infty$
the correlation functions $s(m),z(m) \rightarrow 0$ by large-distance factorisation, and therefore
the entropy at large distances saturates to its maximum value
\begin{equation}
    \lim_{m \rightarrow \infty } S_{\{1,m+1\}}=2\log(2),
\end{equation}
which is exactly twice the entanglement entropy of a single spin
computed above, as expected.  One can now evaluate the
entanglement entropy of two spins for a great variety of
spin-$\frac{1}{2}$ models for which the correlation functions are
known.

\begin{figure}[h!]
\begin{center}
\includegraphics[width=7.9cm,height=6cm,angle=0]{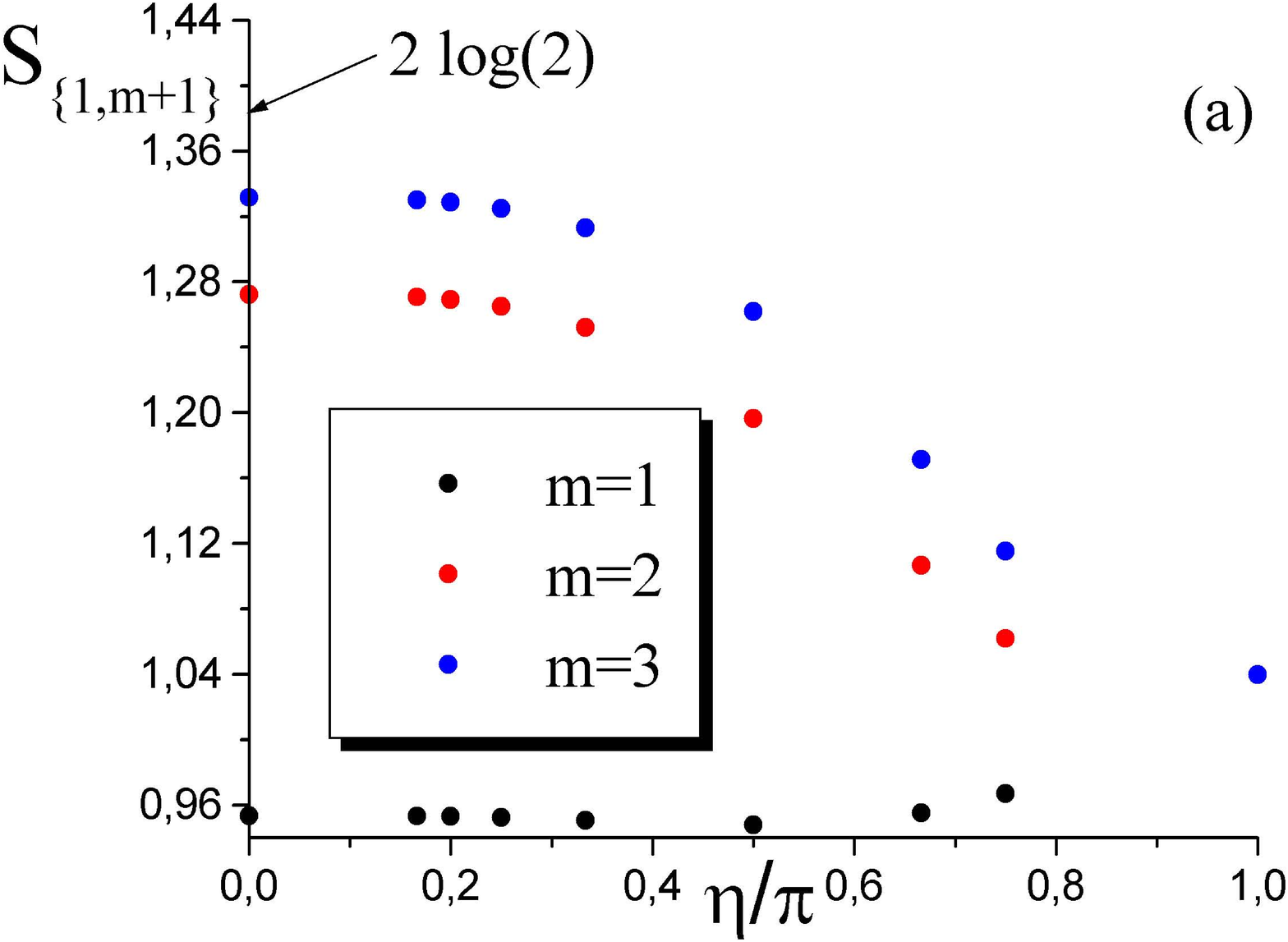}
\includegraphics[width=7.9cm,height=6cm,angle=0]{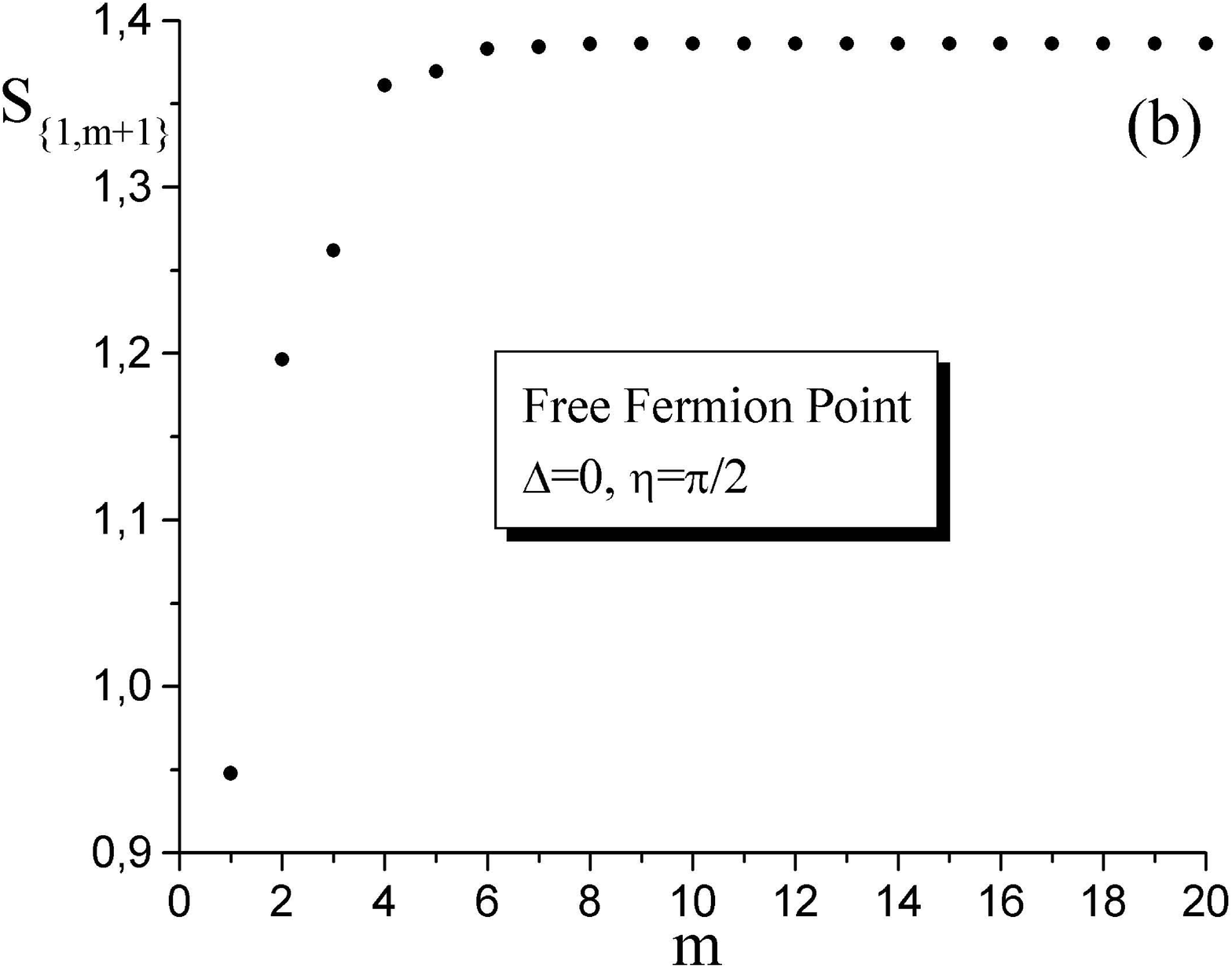}
\caption{(a) The entanglement entropy of two spins $S_{\{1,m+1\}}$
in the XXZ chain for different values of the anisotropy parameter
and $m=1,2,3$. We observe how the entropy tends to grow with $m$
until reaching the value $2\log(2)$ for very large $m$. An
exception to this is the case $\eta=\pi$. In this case the entropy
takes the value $\frac{3}{2}\log(2)$ irrespectively of $m$. This
is due to the very special
 nature of this point, for which the two point function $s(m)$
 oscillates between the values $\pm \frac{1}{4}$ as $m
\rightarrow \infty$ (rather than vanishing). The limit $\eta
\rightarrow \pi^-$ or $\Delta\rightarrow -1^+$ will be discussed
at length in Section \ref{seclimit}. (b) The entanglement entropy
of two spins $S_{\{1,m+1\}}$ at the Free Fermion point, where the
correlation functions (see e.g. \cite{JM4} for a derivation) are
given by: $
    \langle \sigma^{z}_1
    \sigma^{z}_{m+1}\rangle=\frac{2((-1)^m-1)}{\pi^2 m^2}
$
and
$
    \langle \sigma^{+}_1
    \sigma^{-}_{m+1}\rangle=\frac{(-1)^m}{2}\prod\limits_{k=1}^{[\frac{m}{2}]}
    \frac{\Gamma(k)^2}{\Gamma(k-\frac{1}{2})\Gamma(k+\frac{1}{2})}
    \prod\limits_{k=1}^{[\frac{m+1}{2}]}\frac{\Gamma(k)^2}{\Gamma(k-\frac{1}{2})
    \Gamma(k+\frac{1}{2})}.
$ }\label{fig2}
 \end{center}
\end{figure}

We will  conclude this subsection by providing some numerical
results for the entanglement entropy $S_{\{1,m+1\}}$ at particular
values of $m$ and $\Delta$. Although the general expressions for
the two-point functions involved in (\ref{re}) are given in
general in terms of complicated multiple integral representations
\cite{JMi3,JMi,JMi2,ss-maillet}, those integrals have been done at
least for small values of $m$ and different values of the
anisotropy parameter. Very useful tables listing explicit values
of the two-point functions for many values of $\Delta$ and $m=1,
2$ and 3 can be found in \cite{Kato:2003qb,Kato:2004ww}. Employing
those results, it is possible to evaluate the entropy for various
values of the anisotropy parameter as shown in Fig.~\ref{fig2}(a).
In addition, the two-point functions of spin operators admit a
simple form for special values of the anisotropy parameter. One
such value is $\Delta=0$ or $\eta=\frac{\pi}{2}$ which is commonly
known as the free Fermion point (see Fig.~\ref{fig2}(b)).

Finally, we note that the correction terms for the asymptotic
behaviour of the entanglement entropy at large $m$ and generic
values of the anisotropy parameter can be obtained by employing
the results of \cite{Lukyanov:2002fg}. In this paper, the
asymptotic behaviour of the two-point functions $\langle
\sigma^z_1 \sigma^z_{m+1}\rangle$ and $\langle \sigma^x_1
\sigma^x_{m+1}\rangle$ was obtained using perturbed CFT techniques
(the results of \cite{Lukyanov:2002fg} were even more general, as
they considered time-dependent correlation functions).

Using
\begin{equation}\label{A}
    A =\frac{1}{2(1-{\frac{\eta}{\pi}})^2}\left(\frac{\Gamma\left(\frac{{{\eta}}}{2\pi-2{{\eta}}} \right)}{2\sqrt{\pi}\Gamma\left( \frac{\pi}{2\pi-2{{\eta}}}\right)}
    \right)^{\frac{\eta}{\pi}} \exp\left[-\int_{0}^\infty \frac{dt}{t}\left(\frac{\sinh({\frac{\eta}{\pi}} t)}{\sinh t \cosh (1-{\frac{\eta}{\pi}})t}-{\frac{\eta}{\pi}} e^{-2t}
    \right)\right],
\end{equation}
we obtain the following leading asymptotic expression for the
entropy:
\begin{equation}
    S_{\{1,m+1\}}\sim 2 \log(2)-\frac{16 A^2}{m^{{\frac{2\eta}{\pi}}}} -\frac{512 A^4}{8
    m^{{\frac{4\eta}{\pi}}}}+
    \mathcal{O}(m^{-{\frac{6\eta}{\pi}}},m^{-\frc\eta\pi-\frc{4\pi}\eta+4},m^{-\frc\pi\eta},m^{-2}),
\end{equation}
for $\eta \in (0,\pi)$.

\section{Quantum states for chains of infinite length} \label{ssectTop}

Infinite-length chains can be described using Hilbert spaces.
However, contrarily to the case of finite-length chains, the
infinite-dimensional Hilbert spaces ${\cal H}_\Delta$ occurring in
the limit $N\to\infty$ depend, in a sense, on the parameter
$\Delta$ characterising the Hamiltonian (\ref{h})\footnote{Of
course, as a Hilbert space, the space with an infinite countable
basis is unique.}. More precisely, ${\cal H}_\Delta$ is formed by
the completion of the set of finite-energy eigenvectors of the
Hamiltonian $H_\Delta$, and these are sets of very different
vectors for different values of $\Delta$. In particular, the
Hilbert spaces ${\cal H}_\Delta ,\ {\cal H}_{\Delta'}$ are
orthogonal for $\Delta\neq \Delta'$; e.g. in the limit of infinite
volume, the overlap ${}_\Delta\bra
\psi|\psi\ket_{\Delta'}/\sqrt{{}_\Delta\bra
\psi|\psi\ket_{\Delta}{}_{\Delta'}\bra \psi|\psi\ket_{\Delta'}}$
between the two ground states tend to zero. Due to this, the
description of infinite-length quantum chains via Hilbert spaces
is not very well adapted to the study of quantum averages of
localised observables. Indeed, the quantum average of an operator
that factorises to the identity on all sites except finitely many
is well defined on the Hilbert spaces ${\cal H}_\Delta$ for any
$\Delta$ (in the range $-1<\Delta\leq 1$ considered), and smooth
as a function of $\Delta$. The Hilbert space description does not
provide any natural topology in agreement with continuity in
$\Delta$. The study of the limit $\Delta\to-1^+$ requires a more
adapted description.

For Hilbert spaces of finite dimension, knowing all quantum averages uniquely fix the quantum state. From this, one way of describing quantum states (pure or mixed) that is more convenient when considering the infinite-volume limit, is through linear functionals representing quantum averages. Let us consider the infinite set of vector spaces $\{V_i:i\in\Z\}$: this is the set of sites of the chain as usual, except that the index can be both positive and negative (i.e. there is no boundary). For any finite block of sites $A$, let us consider the complex vector space ${\cal E}_A$ of operators supported on $A$: the space $\otimes_{i\in A} {\rm End}(V_i)$ with action like the identity on all sites $j\not\in A$. Let us further consider the complex vector space ${\cal E}=\cup_A {\cal E}_A$ of finitely-supported operators; explicitly, these are the finite linear combinations of finite products of elementary matrices $E_{i}^{\ep\ep'}$.

\begin{defi}
The space of quantum states on the infinite-length chain, which we will denote by ${\cal F}$, is the space of all linear functionals $\psi:{\cal E}\to \C$ that are real-linear:
\beq
    \psi(E)^* = \psi(E^\dag)\quad (E\in{\cal E})
\eeq
and normalised
\beq
    \psi(1) = 1.
\eeq
\end{defi}
The physical meaning of $\psi(E)$, for hermitian $E$, is simply the average of the observable $E$ in the physical state represented by $\psi$.

For infinite length, this definition of a quantum state is extremely useful, because it puts in a common space the quantum states of any quantum chain with well-defined correlation functions: for instance, the XXZ chains for $-1<\Delta\leq 1$. Indeed, to every ray $\{a|\psi\ket:a\in\C\}$ lying in the Hilbert space ${\cal H}_\Delta$ for some $\Delta$, we associate a quantum state in ${\cal F}$ defined by
\beq\label{linf}
    \psi(E) := \frc{\bra \psi| E |\psi\ket}{\bra\psi|\psi\ket},\quad E\in{\cal E}.
\eeq

This definition of the space of quantum states is also very
natural from the viewpoint of the R\'enyi entropies
$S_A^{\text{R\'enyi}}(n)$ for integer $n$. Given a quantum state
$\psi\in{\cal F}$, we can construct the linear functional
$\Psi:=\psi^{\otimes n}$ on ${\cal E}^{\otimes n}$, and the R\'enyi
entropy of $\psi$ with respect to a finite set of sites $A$ is
expressed using our replica permutation operators:
\beq\label{SAstate}
    S_A^{\text{R\'enyi}}(n) = \frc1{1-n}\log\lt(\Psi\lt({\cal T}_A\rt)\rt).
\eeq Note that indeed, the permutation operator ${\cal T}_A$
associated with a finite block of sites $A$ is a
finitely-supported operator. Naturally, since the density matrix
$\rho_A$ is completely determined by the fundamental blocks on the
sites $A$, it can also be defined in the infinite-volume case. In
particular, we have \beq
    \Tr\lt(\rho_A^n\rt) = \Psi\lt({\cal T}_A\rt).
\eeq

Recall the concept of local-unitary transformations introduced in
section \ref{sec2}. This concept can naturally be included into
the infinite-length setting. Given disjoint blocks of sites $A_k$,
$k=-\infty,\ldots,\infty$, a {\em block-defined transformation} is
a doubly-infinite sequence $U=(U_{k}:k\in\Z)$ of elements
$U_{k}\in {\cal E}_{A_k}$ that are invertible,
$U_{k}^{-1}\exists\;\forall\;k$. To any such $U$ we can associate
an adjoint action on ${\cal E}$, given by ${\rm Ad}\; U (E) :=
\prod_k U_k^{-1}EU_k$, $E\in{\cal E}$. The product is finite,
hence this is a well-defined action. This gives rise to an action
on quantum states, defined by $U\cdot \psi = \psi \circ {\rm Ad}\;
U$ for any $\psi\in{\cal F}$. A local-unitary transformation is a
block-defined transformation, where $A_k = \{k\}$ all $U_{k}$ are
unitary. It stays true in the infinite-length setting that ${\cal
T}_A$ is local-unitarily invariant.

In order to discuss convergence of  quantum states in ${\cal F}$,
we define a topology (in fact, a geometry), via compact
convergence of fundamental blocks. This topology in fact works
(and will be used) more generally for linear functionals without
the normalisation condition ($\psi(1)=1$). This is inspired by the
usual compact convergence topology in function theory. Consider a
sequence of functions on ${\cal F}$ parametrised by an integer
$k\ge 0$, defined by
\[
    d_k(\psi_1,\psi_2) := \sqrt{\sum_{\{\ep_i,\ep_i':i=-k,\ldots,k\}}
        \lt|\psi_1\lt(\prod_{i=-k}^k E_i^{\ep_i\ep_i'}\rt) -
            \psi_2\lt(\prod_{i=-k}^k E_i^{\ep_i\ep_i'}\rt) \rt|^2}
                \quad (\psi_1,\; \psi_2\;\in {\cal F}).
\]
This is a natural definition, because it is invariant under a
change of orthonormal basis in the space of linear operators on
the subchain of $2k+1$ sites centered at 0 (with the inner product
given by $(E,E') = \Tr(E^\dag E')$). Consider then the distance function
\beq\label{dist}
    d(\psi_1,\psi_2) := \sum_{k=0}^\infty \frc{2^{-k} d_k(\psi_1,\psi_2)}{1+d_k(\psi_1,\psi_2)}.
\eeq

\begin{defi}
The compact convergence topology on the space of quantum states
${\cal F}$ is that of open balls induced from the distance
function (\ref{dist}).
\end{defi}

This topology is the most convenient when considering
finitely-supported observables. Indeed, by the usual arguments of
compact convergence, one can see that every Cauchy sequence
$(\psi_j\in {\cal F}:j=1,2,\ldots)$ gives rise, for any
finitely-supported operator $E$, to a Cauchy sequence
$(\psi_j(E):j=1,2,\ldots)$. In other words, the limit of a Cauchy
sequence in ${\cal F}$ is in ${\cal F}$: the space of quantum
states is complete. Moreover, again by standard arguments of
compact convergence, if, for a sequence $(\psi_j\in {\cal
F}:j=1,2,\ldots)$, all fundamental blocks
$\psi_j\lt(\prod_{i=m_1}^{m_2} E_i^{\ep_i\ep_i'}\rt)$ converge as
$j\to\infty$, then the sequence converges, and the limit is a
quantum state whose action on elementary matrices agrees with the
limit of the fundamental blocks. That is, this is the topology
that associate convergence of quantum states to convergence of
averages of all finitely-extended observables.

We remark that this topology also allows us to describe the infinite-length limit of quantum chains. Indeed, to any sequence of vectors $|\psi\ket^{(k)}$ in the space spanned by the sites from $-k$ to $k$, for $k=1,2,3,4,\ldots$, we can associate a sequence $\psi_k$ of elements in ${\cal F}$, by $\psi_k(E) = {}^{(k)}\bra\psi|E|\psi\ket^{(k)}$ if $E\in {\cal E}_{\{-k,\ldots,k\}}$, and $\psi_k(E)=0$ otherwise. If $|\psi\ket^{(k)}$ describe ground states, and if the infinite-volume limit of the chain exists, then this sequence converges, hence define a state $\lim_{k\to\infty}\psi_k\in {\cal F}$. The same procedure can be applied for any sequence of linear functionals on spaces ${\cal E}_{A_k}$ with growing $A_k$ such that $\lim_{k\to\infty} A_k = \Z$. Let us also note that in this context, there is no {\em a priori} clear distinction between pure states and mixed states in infinite volume. Indeed, for any density matrix $\rho$ on a finite number of sites, it is always possible to construct a pure state on twice as many sites, such that the reduced density matrix equals $\rho$. Hence, to a sequence of density matrices leading to a converging sequence of quantum states, we can associate a sequence of vectors giving rise to the same limit quantum state (and vice versa).

\section{The limit $\Delta\to-1^+$}\label{seclimit}

We now wish to study the state occurring in the limit
$\Delta\to-1^+$ of the infinite-length XXZ model in compact
convergence topology. Using the calculation Rules \ref{4rules}, as
well as exact results for fundamental blocks of the XXZ models
found in the literature, we provide a full description of the
reduced density matrix associated to any finite subset of sites.
Our main result is that the reduced density matrix associated with
a number $m$ of sites (no matter what their positions are) has
only $m+1$ non-vanishing eigenvalues which are given by \beq
\lambda_k=\frac{1}{2^m} \left(\begin{array}{c}
  m \\
  k\\
\end{array}\right) \qquad \text{with} \qquad k=0,\ldots, m.
\label{eigenden}
\eeq
Therefore, for $m$ odd each eigenvalue is twice degenerate,
whereas for $m$ even all eigenvalues but one are twice degenerate.

\subsection{The XXZ quantum state in the limit $\Delta\to-1^+$}

From the explicit formulas obtained in
\cite{JMi3,JMi,JMi2,ss-maillet}, it follows that the limit $\Delta
\to -1^+$ of the fundamental blocks associated with the ground
state of the Hamiltonian (\ref{h}) exists and is finite. More
precisely, specializing the multiple integral formulae obtained in
\cite{JMi3,JMi,JMi2,ss-maillet}, we find that in the limit
$\Delta\to-1^+$, the non-vanishing fundamental blocks associated
to an index set $A=\{j_1,j_2,\ldots,j_m\}$ (with $j_k\neq j_{k'}$
for $k\neq k'$) reduce to
\begin{equation}
   \lim_{\Delta\to-1^+} \langle E^{\epsilon_1 \epsilon'_1}_{j_1} E^{\epsilon_2 \epsilon'_2}_{j_2}
   \cdots E^{\epsilon_m \epsilon'_m}_{j_m}\rangle=\frac{1}{2^m} \prod_{j \in
   B}(-1)^j ,\label{pro}
\end{equation}
where $B$ is the subset of sites at which either an operator $E^{12}$ or an operator $E^{21}$ sits:
\[
    B = \{j_k:\ep_k\neq\ep'_k\}\subset A.
\]
Note that the indices $j_1,\ldots,j_m$ in formula (\ref{pro}) are not necessarily consecutive, contrarily to the formula for fundamental blocks in \cite{JMi2,KMT}. It is simple to check that (\ref{pro}) is consistent with the relation ${\bf 1} = E_{j}^{11} + E_j^{22}$.

Let us consider the ground state $|\psi\ket_\Delta$ of the Hamiltonian $H_\Delta$. For $\Delta\in(-1,1]$, the corresponding quantum states $\psi_{\Delta}$ all lie in ${\cal F}$. Moreover, by formula (\ref{pro}), all fundamental blocks converge in the limit where $\Delta$ tends to $-1$ from above. Hence, the limit $\lim_{\Delta\to-1^+}\psi_{\Delta}$ exists and is a quantum state of the infinite-volume chain. We will denote this limit by
\beq
    \psi_{-}:=\lim_{\Delta\to-1^+}\psi_{\Delta}\ \in{\cal F}.
\eeq
This quantum state is completely characterised by formula (\ref{pro}):
\beq\label{blockstate}
    \psi_-\lt(E^{\epsilon_1 \epsilon'_1}_{j_1} E^{\epsilon_2 \epsilon'_2}_{j_2}
   \cdots E^{\epsilon_m \epsilon'_m}_{j_m}\rt) = \lt\{\ba{ll} \frac{1}{2^m} \prod_{j \in
   B}(-1)^j & \text{equal numbers of $E^{12}$ and $E^{21}$} \\ 0 & \text{otherwise.} \ea\rt.
\eeq

We would now like to study this state by computing its R\'enyi and von Neumann entanglement entropies. As is clear from (\ref{blockstate}), a very special feature of this state is that its correlation functions are extremely simple. Hence, we will be able to find analytic expressions for the R\'enyi entropies and deduce the exact eigenvalues and degeneracies of the reduced density matrices; these are inaccessible for generic values of $\Delta$.

\subsection{R\'enyi and von Neumann entanglement entropies of the state $\psi_{-}$} \label{ssectRenyi}

The R\'enyi entropy (\ref{renyi}) can be evaluated using the replica permutation operators via (\ref{SAstate}). The present subsection provides a proof of the following proposition, which is in agreement with the first point stated at the beginning of this section, equation (\ref{eigenden}).
\begin{propo}\label{propoR}
Consider the state $\psi_-$, obtained in the limit $\Delta\to-1^+$
of the ground state $\psi_\Delta$ in the compact convergence
topology. Its R\'enyi and von Neumann entanglement entropies
associated with any set of $m$ sites are given by
\beq\label{renyi4}
  S^{\text{\rm R\'enyi}}_A(n)=-\frac{n m \log 2}{1-n} + \frac{1}{1-n}\log\left(\sum_{k=0}^m\left(\begin{array}{c}
  m\\
  k\\
\end{array}\right)^n\right)
\eeq
and
\begin{equation}\label{vonneu2}
  S_A= m \log 2 - \frac{1}{2^{m}}\sum_{k=0}^{m}
 \left(\begin{array}{c}
  m\\
  k\\
\end{array}\right) \log
 \left(\begin{array}{c}
  m\\
  k\\
\end{array}\right).
\end{equation}
The large-$m$ asymptotics of these quantities are given by
\beq
    S^{\text{\rm R\'enyi}}_A(n) = \frc12\log\lt(\frc{\pi m}2\rt) + \frc{\log(n)}{2(n-1)} + O\lt(m^{-1}\rt),\quad
    S_A = \frc12\log\lt(\frc{\pi m}2\rt) + \frc12 +
    O\lt(m^{-1}\rt).\label{asims}
\eeq The leading term in the asymptotics of the R\'enyi entropy is
therefore independent of $n$.
\end{propo}

Let us start with the simplest non-trivial case $n=2$. The specialisation of (\ref{pitu}) to $n=2$ immediately gives
the formula
\begin{equation}
   \Tr\lt(\rho_A^2\rt)=\sum_{\ep_i',\epsilon_i=1}^2 \psi_-\lt(E^{\epsilon_1 \epsilon'_1}_{j_1} E^{\epsilon_2 \epsilon'_2}_{j_2}
   \cdots E^{\epsilon_m \epsilon'_m}_{j_m}\rt)\psi_-\lt(
   E^{\epsilon'_1 \epsilon_1}_{j_1} E^{\epsilon'_2 \epsilon_2}_{j_2}\cdots E^{\epsilon'_m
   \epsilon_m}_{j_m}\rt).\label{rho5}
\end{equation}
Notice that many of the correlation functions in this sum are
vanishing: a fundamental block is non-zero if and only if it involves equal numbers of matrices $E^{12}$ and $E^{21}$.

Given the structure of the product (\ref{rho5}) it is easy to see
that the products of $(-1)^j$ in (\ref{pro}) will not play a role,
as matrices $E^{12}$ and $E^{21}$ appear at exactly the same sites
in both correlation functions, so that all phase factors cancel
out. Therefore, the sum (\ref{rho5}) reduces to
\begin{equation}
  \Tr\lt(\rho_A^2\rt)=\frac{\text{number of non-vanishing m-site fundamental blocks}}{2^{2m}}.
\end{equation}
The number of non-vanishing $m$-site blocks can be easily
obtained by simple combinatoric arguments. A
generic non-vanishing fundamental block will have $s$ operators
$E^{11}$, $q$ pairs of operators $E^{12}$ and $E^{21}$, and $m-s-2q$ operators
$E^{22}$ at the remaining sites. The total number of such
blocks is simply
\begin{equation}
    \frac{m!}{s! (q!)^2 (m-s-2q)!},
\end{equation}
therefore
\begin{equation}
    \Tr\lt(\rho_A^2\rt)=\frac{1}{2^{2m}}\sum_{q=0}^{[\frac{m}{2}]}
    \sum_{s=0}^{m-2q} \frac{m!}{s! (q!)^2
    (m-s-2q)!}=\frac{\Gamma\left(\frac{1}{2}+m\right)}{\sqrt{\pi}\Gamma\left(1+m\right)}.
\end{equation}
The R\'enyi entropy (\ref{renyi}) is given by
\begin{equation}
    S^{\text{R\'enyi}}_m (2)
    =-\log\left(\frac{\Gamma\left(\frac{1}{2}+m\right)}{\sqrt{\pi}\Gamma\left(1+m\right)}\right),
\end{equation}
and, for $m$ large behaves asymptotically as
\begin{equation}\label{mlarge}
    S^{\text{R\'enyi}}_m (2) = \frac{1}{2}\log(\pi m) +
    O(m^{-1}).
\end{equation}

The calculation can be extended to higher values of $n$. The R\'enyi entropy for generic $n$ will be obtained form a sum over an
$n$-fold product of $m$-site fundamental blocks. A feature of Rules \ref{4rules} is that for every site, there must be as many matrices $E^{12}$ as there are matrices $E^{21}$ distributed amongst the $n$ copies. Hence, as in the $n=2$ case, the $(-1)^j$ factors in (\ref{pro}) all cancel out when multiplying the $n$ fundamental blocks together, because any possible factor $-1$ will always be present an even number of times. This means that once more, the computation reduces to combinatorially working out the number of non-vanishing products of fundamental blocks that occur.

A simplification to this combinatorial calculation is obtained
from two general consequences of Rules \ref{4rules}. First, once a
choice of elementary matrices has been made for the first $n-1$
fundamental blocks, the choice for the $n^{\text{th}}$ one exists
and is unique. This is because the rules say if, on any given
site, we have the elementary matrices $E^{\ep_+\ep_+'}$ at copy
$\alpha+1$ and $E^{\ep_-\ep_-'}$ at copy $\alpha-1$, then the
elementary matrix at copy $\alpha$ is fixed to be
$E^{\ep_+'\ep_-}$. Second, if the first $n-1$ fundamental blocks
are non-vanishing and satisfy the rules, then the unique choice of
the last block also is non-vanishing. This is because there is the
same number of matrices $E^{12}$ as there is of $E^{21}$ in total
in the first $n-1$ blocks (since this is true in every block).
Since at every site there can only be, in the $n-1$ first copies,
an equal number of $E^{12}$ and $E^{21}$ or a surplus by one of
either $E^{12}$ or $E^{21}$, then the number of sites where there
is a surplus of $E^{12}$ must be the same as that where there is a
surplus of $E^{21}$. Hence, Rules \ref{4rules} imply that on the
last block, there will be the same numbers of $E^{12}$ and
$E^{21}$. Therefore, putting these three observations together, in
order to count the number of non-vanishing products of $n$
fundamental blocks occurring, we only need to count the number of
choices of elementary matrices for the first $n-1$ fundamental
blocks that make these blocks non-vanishing and that satisfy Rules
\ref{4rules}.

Let us now analyse the case $n=3$, where the quantity $\Tr\lt(\rho_A^3\rt)$ is given by a sum over triple products of
$m$-site fundamental blocks. We can depict Rules \ref{4rules} through diagrams such as
\begin{center}
\begin{tabular}{c ccccccccc}
  \text{1st fundamental block} && $E^{11}_j$ &  & & &  & $E^{12}_j$ &  &  \\
 & $\swarrow$  & &$\searrow$ & & & $\swarrow$  & &$\searrow$ &\\
 \text{2nd fundamental block} &  $E^{11}_j$ & \text{or}  & $E^{21}_j$ & &  & $E^{11}_j$ & \text{or} & $E^{21}_j$ &  \\
  & $\downarrow$ & & $\downarrow$ & & & $\downarrow$ & & $\downarrow$ &\\
   \text{3rd fundamental block} & $E^{11}_j$ &  \text{or} &  $E^{12}_j$ & & & $E^{21}_j$ &  \text{or} &   $E^{22}_j$ & \\
\end{tabular}
\end{center}
and similarly for $E^{22}_j$ and $E^{21}_j$ by exchanging
super-indices $1,2$ everywhere.

Let us consider the first fundamental block. Let it contain $s$
matrices $E^{11}$, $q$ pairs of matrices $E^{12}, E^{21}$, and $m-s-2q$
matrices $E^{22}$. Each such fundamental block can appear in
conjunction with various choices of the second fundamental block. According to the rules above, every matrix $E^{11}$ or
$E^{12}$ in the first can be followed by a matrix
$E^{11}$ or $E^{21}$ in the second.
Similarly, every matrix $E^{22}$ or $E^{21}$ in the first
can be followed by a matrix $E^{22}$ or $E^{12}$ in the
second. The number of second fundamental blocks containing $a$ pairs of matrices $E^{12}$ and $E^{21}$ is given by
\begin{equation}\label{numblocks}
    \left(\begin{array}{c}
  s+q \\
  a\\
\end{array}\right)  \left(\begin{array}{c}
  m-s-q \\
  a\\
\end{array}\right).
\end{equation}
Summing over all possible values of $a$ gives
\begin{equation}\label{sumnumblocks}
    \sum_{a=0}^{\min(s+q,m-s-q)}\left(\begin{array}{c}
  s+q \\
  a\\
\end{array}\right)  \left(\begin{array}{c}
  m-s-q \\
  a\\
\end{array}\right)= \left(\begin{array}{c}
  m \\
  s+q\\
\end{array}\right).
\end{equation}

Therefore, the total number of distinct triple products leads to
\begin{eqnarray}
    \Tr \left( \rho^3_A \right) &=&\frac{1}{2^{3m}}\sum_{q=0}^{[\frac{m}{2}]}
    \sum_{s=0}^{m-2q} \frac{m!}{s! (q!)^2
    (m-s-2q)!}\left(\begin{array}{c}
  m\\
  s+q\\
\end{array}\right) \nonumber\\
    &=&\frac{\Gamma\left(\frac{1}{2}+m\right)}{2^m
    \sqrt{\pi}\Gamma\left(1+m\right)}\,
    {}_3F_2\left(\left\{\frac{1}{2}-m,-m,-\frac{m}{2}\right\}
    ,\left\{1,\frac{1}{2}-m\right\};1\right).\label{rho3}
\end{eqnarray}
The leading asymptotic behaviour of the R\'enyi entropy can be
extracted by exploiting the fact that for large $m$
\begin{eqnarray}
   && {}_3F_2\left(\left\{\frac{1}{2}-m,-m,-\frac{m}{2}\right\},\left\{1,\frac{1}{2}-m\right\};1\right)
   \sim
   {}_3F_2\left(\left\{-m,-m,-\frac{m}{2}\right\},\left\{1,-m\right\};1\right)\nonumber\\
   &&=
   {}_2F_1\left(\left\{-\frac{m}{2},-\frac{m}{2}\right\},\left\{1\right\};1\right)=
   \frac{m!}{\Gamma\left(1+\frac{m}{2}\right)^2}.
\end{eqnarray}
Substituting this into (\ref{rho3}) and employing Stirling's
approximation for the Gamma functions we obtain,
\begin{equation}
   S^{\text{R\'enyi}}_m (3) = \frac{1}{2} \log(\frac{\pi m}{2})+
   O(1,m^{-1}).
\end{equation}
Unfortunately, the approximation above is too crude and does not
allow to extract the correct constant $1/4 \log(3)$ as predicted
by (\ref{asims}). This is the same leading behaviour as in
(\ref{mlarge}) which is incompatible with the CFT prediction
according to which the coefficient of $\log(m)$ should be a
function of $n$.

It is possible to extend the previous combinatorial arguments to higher (generic) values of $n$. Again, from Rules \ref{4rules}, we know that it is from the set of matrices $E^{11}$ and $E^{21}$ in the fundamental block of copy $\alpha$ that emerge all matrices $E^{11}$ and $E^{12}$ in the copy $\alpha+1$. The same is true if all super-indices 1,2 are
exchanged. Let us consider the $\alpha^{\text{th}}$ fundamental block, for $\alpha=1,2,\ldots,n-1$. Let it contain $s_\alpha$
matrices $E^{11}$, $q_\alpha$ pairs of matrices $E^{12}, E^{21}$, and $m-s_\alpha-2q_\alpha$
matrices $E^{22}$ (with $s:=s_1$ and $q:=s_1$). Since $q_{\alpha+1}$ matrices $E^{21}$ are chosen from the $s_\alpha+q_\alpha$ matrices $E^{11}$ and $E^{12}$, the rest must be all matrices $E^{11}$, so we must have
\beq
    s_{\alpha+1} = s_\alpha+q_\alpha-q_{\alpha+1}.
\eeq
Hence, we find
\beq
    s_{\alpha+1}+q_{\alpha+1} = s_\alpha+q_\alpha\;\Rightarrow\; s_{\alpha}+q_\alpha = s+q\;\forall\;\alpha.
\eeq
Therefore, from expression (\ref{numblocks}), the number of fundamental blocks at copy $\alpha$ is given by
\begin{equation}
    \left(\begin{array}{c}
  s+q \\
  q_\alpha\\
\end{array}\right)  \left(\begin{array}{c}
  m-s-q \\
  q_\alpha\\
\end{array}\right).
\end{equation}
Summing over all possible values of $q_{\alpha}$ for $\alpha=2,\ldots,n-1$ and using formula (\ref{sumnumblocks}) gives
\begin{equation}
    \Tr\lt(\rho_A^n\rt)=\frac{1}{2^{n m}}\sum_{q=0}^{[\frac{m}{2}]}
    \sum_{s=0}^{m-2q} \frac{m!}{s! (q!)^2
    (m-s-2q)!}\left(\begin{array}{c}
  m\\
  s+q\\
\end{array}\right) ^{n-2}.
    \label{renyi2}
\end{equation}

From the definition (\ref{renyi}) it is clear that the formula
above should give exactly the sum over all $n$-powers of the
eigenvalues of the density matrix associated to the state
$\psi^-$. However, this is not obvious from (\ref{renyi2}) in its
present form. In order to make this structure clearer it is useful
to rewrite (\ref{renyi2}) in a slightly different way. We will
introduce the new variable $k=s+q$ and write the binomial
coefficient to the power $n-2$ as a product of powers $n$ and
$-2$. The sum becomes,
\begin{equation}
    \Tr\lt(\rho_A^n\rt)=\frac{1}{2^{n m}}\sum_{q=0}^{[\frac{m}{2}]}
    \sum_{k=q}^{m-q} \frac{(k!(m-k)!)^2}{(k-q)! (q!)^2
    (m-k-q)!m!} \left(\begin{array}{c}
  m\\
  k\\
\end{array}\right)^{n}.
\end{equation}
Exchanging the two sums we find the equivalent expression:
\begin{equation}
    \Tr\lt(\rho_A^n\rt)=\frac{1}{2^{n m}}\sum_{k=0}^{m}
    \sum_{q=0}^{m-k} \frac{(k!(m-k)!)^2}{(k-q)! (q!)^2
    (m-k-q)!m!} \left(\begin{array}{c}
  m\\
  k\\
\end{array}\right)^{n}=\frac{1}{2^{n m}}\sum_{k=0}^{m}
 \left(\begin{array}{c}
  m\\
  k\\
\end{array}\right)^{n},
\label{renyi3}
\end{equation}
where the last equality follows from the remarkable identity
\begin{equation}
    \sum_{q=0}^{m-k} \frac{(k!(m-k)!)^2}{(k-q)! (q!)^2
    (m-k-q)!m!} =1.
\end{equation}
The result (\ref{renyi3}) clearly identifies the non-vanishing
eigenvalues of the density matrix as (\ref{eigenden}) and provides
a much simpler expression for the von Neumann and R\'enyi
entropies, given by (\ref{renyi4}) and (\ref{vonneu2}).

Even though we have not been able to carry out the sums
explicitly, it is possible to extract the leading behaviour of
(\ref{renyi4}) and (\ref{vonneu2}) for large $m$. First, we transform the last sum in (\ref{renyi3})
into a multiple integral using a generating function procedure:
\beqa
    \Tr\lt(\rho_A^n\rt) &=& \frc1{2^{mn}}
    \frc1{(2\pi i)^{n-1}}\oint_0
        \prod_{j=1}^{n-1}\frc{dx_j}{x_j}\; \prod_{j=1}^{n-1}(1+x_j)^m\;
        \lt(1+\prod_{j=1}^{n-1}x_j^{-1}\rt)^m\n
        &=& \frc1{(2\pi)^{n-1}} \int_{-\pi}^{\pi} \prod_{j=1}^{n-1} d\theta_j\;
        \prod_{j=1}^{n-1}\cos^m\lt(\frc{\theta_j}2\rt)\;
        \cos^m\lt(\sum_{j=1}^{n-1}\frc{\theta_j}2\rt).
\eeqa
Then, we evaluate the large-$m$ asymptotic using a saddle-point analysis. The main contribution to the integral at large-$m$ will be obtained when the integrand is evaluated at $\theta_j\sim 0$ for all $j$. The leading asymptotics is evaluated by using $\cos(\theta/2)=e^{-\theta^2/2+O(\theta^4)}$:
\beqa\label{typeint}
    \Tr\lt(\rho_A^n\rt) &\sim& \frc1{(2\pi)^{n-1}} \int_{-\infty}^{\infty} \prod_{j=1}^{n-1} d\theta_j\;
        \exp\lt[-\frc{m}8\lt(\sum_{j=1}^{n-1} \theta_j^2 +\lt(\sum_{j=1}^{n-1}\theta_j\rt)^2\rt)\rt] \\
        &=& \frc1{(2\pi)^{n-1}} \int_{-\infty}^{\infty} \prod_{j=1}^{n-1} d\theta_j\;
        \exp\lt[-\frc{m}8\lt(n\theta_1^2 + \sum_{j=2}^{n-1} \theta_j^2\rt)\rt] \n
        &=& \frc1{\sqrt{n}}\lt(\frc{2}{\pi m}\rt)^{\frc{n-1}2}.
\eeqa
Corrections coming from the $O(\theta^4)$ terms in the exponential give rise to a factor $1+O(m^{-1})$.

\subsection{Discussion}

In order to discuss the results of Proposition \ref{propoR}, we
construct the state $\psi_-$ in a different way. Let
$\mathbb{P}_z$ be the operation, on the space of
finitely-supported operators, that projects onto those keeping the
$z$ component of the spin: \beqa
    \mathbb{P}_z:\qquad\qquad {\cal E} &\to& {\cal E}\quad\mbox{(linearly)}\n
        \prod_{j\in A} E_j^{\ep_j,\ep'_j} &\mapsto& \delta_{0,\sum_{j\in A}(\ep_j-\ep_j')} \prod_{j\in A} E_j^{\ep_j,\ep'_j}.
\eeqa
Further, let $U$ be the following local-unitary operator:
\beq\label{Utrans}
    U = \prod_{j\in 2\Z+1} \sigma^z_j,
\eeq
and let $\psi_{\h{e}_x}$ be the following quantum state:
\beq\label{psiv}
    \psi_{\h{e}_x} = \bigotimes_{j\in\Z} \psi_{\h{e}_x;j},
\eeq where $\psi_{\h{e}_x;j}:{\cal E}_{\{j\}}\to \C$ is the
average, on site $j$, obtained from the vector
$|\psi_{\h{e}_x}\ket_j:=(|\uparrow\ket_j +
|\downarrow\ket_j)/\sqrt{2}$ representing a spin pointing in the
positive $x$ direction (this is the eigenvector of $\sigma^x_j$
with eigenvalue 1). Note that the right-hand side of (\ref{psiv})
is a converging doubly-infinite tensor product in the compact
convergence topology. Then, it is a simple matter to see that
\beq\label{psim}
    \psi_- = U\cdot \psi_{\h{e}_x} \mathbb{P}_z.
\eeq
Indeed, from (\ref{psiv}) we notice that $\psi_{\h{e}_x;j}\lt(E_j^{\ep,\ep'}\rt) = 1/2$ for all $\ep,\ep'$. Using that ${\rm Ad}\;U \lt(E_j^{\ep,\ep'}\rt) =  (-1)^j E_j^{\ep,\ep'}$ for $\ep\neq\ep'$, and that ${\rm Ad}\;U \lt(E_j^{\ep,\ep'}\rt) =  E_j^{\ep,\ep'}$ for $\ep=\ep'$, we immediately find that the right-hand side of (\ref{psim}) reproduces formula (\ref{blockstate}). Note that in (\ref{psim}), the order of the operations $U\cdot$ and $\mathbb{P}_z$ is irrelevant, since they commute (i.e. $[{\rm Ad}\;U\,,\,\mathbb{P}_z]=0$).

The form (\ref{psim}) of the state $\psi_-$ points to the following construction.
For each $k=0,1,2,\ldots$, consider the Hilbert space $\otimes_{i=-k}^k V_i$, and the vectors where all spins point in the positive $x$ direction: $|\psi_{\h{e}_x}\ket^{(2k+1)} := \prod_{i=-k}^k |\psi_{\h{e}_x}\ket_i$. Of course, these vectors can be written as linear combinations of vectors with spins pointing in $z$ directions (here, we use the standard values $\pm1/2 \equiv \uparrow,\downarrow$ for the spin variables $s_i$, instead of the numbers $1,2$ respectively):
\[
    |\psi_{\h{e}_x}\ket^{(2k+1)} =
        \frc1{(\sqrt{2})^{2k+1}} \sum_{\{s_i= \pm\frc12:i=-k,\ldots,k\}} \bigotimes_{i=-k}^k |s_i\ket_i.
\]
Now consider the Hilbert space $\otimes_{i=-k}^{k} V_i\otimes
V_{\rm env}$ with $V_{\rm env}\cong \C^{2k+2}$. The space $V_{\rm
env}$ can be though of as a spin-$(k+1/2)$ module, with
orthonormal basis vectors $|S\ket_{\rm env},\;
S\in\{-k-1/2,\ldots,k+1/2\}$ of spins $S/2$. Take the vectors
$|\psi_k\ket$ as follows: \beq\label{const1}
    |\psi_k\ket =
        \frc1{(\sqrt{2})^{2k+1}} \sum_{\{s_i=\pm\frc12:i=-k,\ldots,k\}} \bigotimes_{i=-k}^k |s_i\ket_i
        \otimes \lt|-\sum_{i=-k}^k s_i\rt\ket_{\rm env}.
\eeq
Essentially, we have adjoined to every vector in the sum an additional degree of freedom with a spin exactly opposite to the total spin of the original chain (the sites from $-k$ to $k$), in such a way that all vectors have total spin 0 (in the $z$ direction). In the sub-sum of vectors where the original chain has a given total spin, the adjoined vector just factorises out, so that when calculating averages, the resulting sub-sum of matrix elements is unchanged. Hence, the average in $|\psi_k\ket$ of any product $\prod_{j=-k}^k E_j^{\ep_j,\ep_j'}$ that preserve the total spin of the original chain is exactly equal to the average in $|\psi_{\h{e}_x}\ket^{(2k+1)}$. However, the average of such elementary matrices in $|\psi_k\ket$ is zero if they do not preserve the total spin.

As usual, from $|\psi_k\ket$ we can form a linear functional $\psi_k$ on ${\cal E}_{\{-k,\ldots,k\}}$. By the discussion above, we have
\[
    \psi_k\lt(\prod_{j=-k}^{k} E_j^{\ep_j,\ep_j'}\rt) = \lt\{\ba{ll}
        {}^{(2k+1)}\bra\psi_{\h{e}_x}|\prod_{j=-k}^{k} E_j^{\ep_j,\ep_j'} |\psi_{\h{e}_x}\ket^{(2k+1)} &
            \mbox{total $z$ component preserved} \\
        0 & \mbox{otherwise}.
    \ea\rt.
\]
We may now take the limit
\beq\label{const2}
    \psi_\infty:=\lim_{k\to\infty}\psi_k.
\eeq Since the averages in the states
$|\psi_{\h{e}_x}\ket^{(2k+1)}$ are stable as $k$ increases, and
since the linear functionals act on increasing subsets of the
chain whose limit is the whole chain, we find that the limit
exists in the compact convergence topology. We have $\psi_\infty =
\psi_{\h{e}_x} \; \mathbb{P}_z$, hence \beq\label{const3}
    \psi_- = U\cdot \psi_\infty.
\eeq

The state $\psi_{\h{e}_x}$ is factorisable, hence has entanglement
entropy equal to 0. The higher entanglement entropy of
(\ref{psim}) (in particular, the diverging behaviour at large $m$)
comes from the interplay between this factorised state where spins
point in $x$ directions, and the condition of having a total
$z$-component equal to 0.

Construction (\ref{const1}), (\ref{const2}), (\ref{const3}) gives
some insight into this interplay. It makes it clear that the extra
entanglement entropy generated by the presence of the operator
$\mathbb{P}_z$, i.e. by the condition of preserving the
$z$-component, is due to an extra entanglement with the additional
degree of freedom $V_{\rm env}$. This degree of freedom can be
interpreted as an ``environment'', which couples with the
``system'', the sites from $-k$ to $k$, via the $z$-component of
the total spin. The system can be seen as a random collection of
spins up and down, all configurations having equal probabilities.
Such a state has zero entanglement entropy with respect to any
subsystem. However, the coupling generates entanglement between
the system and the environment, which is picked up when measuring
the entanglement between a subsystem and the rest. Since the
environment is infinite-dimensional in the infinite-$k$ limit, its
capacity for entanglement is infinite, hence as the number of
sites $m$ of the subsystem increases, the entanglement entropy
increases. On the other hand, since the environment does not
couple to each individual spin separately, the entanglement
entropy increases at a rate that is less than linear in $m$. The
number of blocks of system states that couple to separate
environment states is the number of possible values of the
$z$-component of the total spin. For $m$ sites, this is $m+1$,
hence we can indeed expect a logarithmic behaviour in $m$ for the
entanglement entropy at large $m$. Since the blocks have different
sizes, they have different probabilities of occurring, hence we
may indeed expect $b \log m$ for some $b<1$. In fact, the size of
a block with total $z$-spin equal to $k-m/2$, for $k$ between $0$
and $m$, is simply the binomial coefficient $\mato{c} m\\k\matf$.
This explains the eigenvalues of the reduced density matrix, from
which we derived the $(1/2)\log m$ behaviour.

Besides explaining the features of the entanglement entropy of the
state $\psi_-$, this construction also gives an insight into the
way the ground state $\psi_\Delta$ of the XXZ model tends to
$\psi_-$ as $\Delta\to-1^+$. First, note that
$U\cdot\psi_{\h{e}_x}$ is a ground state of the XXZ model at
$\Delta=-1$, because (as is well known) the Hamiltonian of this
model is simply obtained from a $U$ transform of the Hamiltonian
of the {\em ferromagnetic} XXX model -- the ground state energy of
the latter is highly degenerate, and $\psi_{\h{e}_x}$ is a
possible ground state. We expect that as $\Delta\to-1^+$, there is
an increasing characteristic length $\xi$ such that: 1) on scales
much below $\xi$, sites are essentially randomly up and down (in
the $z$-direction), with appropriate coefficients so that they
locally reproduce the ground state $U\cdot\psi_{\h{e}_x}$ of the
XXZ  model at $\Delta=-1$; and 2) on scales above $\xi$, the total
spin is more likely to be zero, as it should be for the XXZ model
for  $-1<\Delta\leq 1$. As $\Delta\to-1^+$, the ``inner'' sites,
on any finite scales, are described by the state
$U\cdot\psi_{\h{e}_x}$, and the  ``outer'' sites, beyond
$\xi\to\infty$, are collectively described by the single degree of
freedom $V_{\rm env}$, whose entanglement with the inner sites,
necessary to keep the total spin to zero, projects onto
spin-preserving operators.

An immediate feature of large-$m$ scaling of the R\'enyi entropy
of the  state $\psi_-$ is that it is in disagreement with the
predictions coming from conformal invariance of critical points.
For instance, for a region $A$ of $m$ consecutive sites, CFT
predicts
\[
    S_A^{\text{R\'enyi}}(n) \stackrel{CFT}= \frc{c}6(1+n^{-1}) \log m + O(1)
\]
where $c$ is the central charge. This is different from
(\ref{asims}), which  is independent of $n$. Yet, the scaling
(\ref{asims}) is logarithmic, hence there is a similar type of
scale invariance (or covariance): a region twice bigger sees its
entanglement entropy added by half a ``unit'', $1/2\log(2)$. The
scale $\xi$ should give rise to a new universal function
interpolating between this non-conformal scale-invariant
behaviour, and the conformal scaling. That is, it should be
possible to define a new scaling limit, looking at the
entanglement entropy for a region of length $m$ that scales like
$m=\alpha \xi$ while the limit $\Delta\to-1^+$ is taken. The
result is likely to be $f(\alpha) \log m$, for a function
$f(\alpha)$ interpolating between $f(0)=1/2$ and the coefficient
predicted by CFT, e.g. $f(\infty) = c/3$ for the von Neumann
entanglement entropy of a connected region, with the central
charge $c=1$.

Another important feature of the entanglement entropy of the state
$\psi_-$ is that it does not depend on the actual positions of the
sites of the region $A$. As will be clear from our future work
\cite{twistagain}, this is a feature of any ground state of the
XXZ model at $\Delta=-1$. It turns out that the state $\psi_-$ is
just one of these ground states. The field theory interpretation
is that the twist field has dimension 0 (corresponding to a
central charge 0), but that there is an entanglement entropy
contribution coming from the degeneracy.

\section{Conclusions and outlook}

In this paper we have described in detail a new approach to the
computation of the bi-partite entanglement entropy of quantum spin
chain models. Our approach employs the ``replica trick" widely
used in the CFT context \cite{Calabrese:2004eu, Calabrese:2005in}
and also, more recently, for integrable QFTs \cite{entropy}. This
approach consists of considering a ``replica" version of the
theory for which one wants to compute the entropy, that is, a new
theory consisting of $n$ non-interacting copies of the original
model. It turns out that computations of the entropy are often
more feasible in the replica model than in the original seemingly
simpler theory. The reason for this is the existence of new
symmetries of the replica model which enable extra fields (in the
QFT case) or extra local operators (in the quantum spin chain
case) to exist in the $n$-copy model. These fields (operators) are
related to the symmetry of the replica theory under cyclic
permutations of the $n$-copies.

For integrable QFTs the fields referred to above were introduced
in \cite{entropy} and named branch point twist fields. It was
shown that the entanglement entropy can be expressed in terms of
their two-point functions which lead to various computation of the
entropy which have been summarized in the review \cite{review}.

In this paper we have identified the basic operators in terms of
which the spin chain equivalent of the QFT twist field can be
expressed: they are \emph{local cyclic replica permutation
operators}. Each operator $\mathcal{T}_i$ acts on one site of the
chain $i$ by cyclicly permuting the spins of the $n$-copies at
that particular site. Ground state correlation functions of
products of such operators over different sites of the chain play
the same role as two (or higher)-point functions of branch point
twist fields in QFT.

We have described a representation of the replica permutation
operators for quantum spin chains of general spin in terms of
fundamental $2 \times 2$ matrices
$E_{\alpha,i}^{\epsilon\epsilon'}$ acting at site $i$, copy
$\alpha$ of the chain and proved that such representation does
reproduce the cyclic permutation action on quantum states
described above. We have also shown that the permutation operators
satisfy exchange relations with respect to other local operators
of the chain, in much the same way as the twist field in QFT. In
fact, as in QFT, these exchange relations can be used as starting
point for the construction of the permutation operator in quantum
spin chains.

We have demonstrated the working of our approach by evaluating the
bi-partite entanglement entropy of one and two spins in an
infinitely long spin-$\frac{1}{2}$ quantum spin chain
characterized by an antiferromagnetic ground state. We find that
the computation of the entropy is recast into combinatorial
problems, which become less tractable as the number of spins is
increased. It will be very interesting to address this problem in
the future and to establish whether or not our approach may be
useful for the study of the entanglement entropy of large
subsystems. From our approach it is clear that the larger the
subsystem the more operators will be involved in the correlation
functions which enter the expressions of the entropy, so that
advancement in the understanding of the large-distance behaviour
of such correlation functions (see \cite{Kitanine:2008bs} for
recent progress) goes hand in hand with the success of our
approach to evaluate the large size asymptotics of the entropy.

Despite the difficulties emphasized above we have been able to
identify a particular quantum state for which both the von Neumann
and R\'enyi entropies are non-trivial but can still be explicitly
computed for any subsystem size and any values of $n$. This
quantum state, which we denoted by $\psi_-$, is defined by its
correlation functions which are obtained as the $\Delta
\rightarrow -1^+$ limit of those of the spin-$\frac{1}{2}$ XXZ
quantum spin chain in the antiferromagnetic regime
\cite{JMi3,JMi,JMi2,ss-maillet}. Our analysis has allowed us to
identify all the eigenvalues of the density matrix associated to
this quantum state and to find the exact large subsystem
asymptotics of both the von Neumann and R\'enyi entropies. In both
cases the leading term scales with the logarithm of the size of
the subsystem as for critical systems. However the coefficient of
this logarithmic term is independent of $n$ for the R\'enyi
entropy in direct contrast with the well-known dependence found
for CFTs.

It is tempting to speculate that this logarithmic but
non-conformal behaviour may be somehow related to the unusual
behaviour at $\Delta=-1$ encountered in \cite{Ercolessi:2010eb}.
In this recent work it is argued that such unusual features are
characteristic of essential singularities. We must however stress
that the quantity evaluated in \cite{Ercolessi:2010eb} and in the
present work are rather different in nature, as well as the
characterization of the quantum state of the chain, so that the
unusual features observed in both cases are not a priori related
in an obvious way. Yet, it is possible that the description that
we gave of how the ground state approaches the $\Delta=-1$ point
from the region $\Delta>-1$ may be of use in understanding the
results of \cite{Ercolessi:2010eb}. For instance, it may be that
the characteristic length $\xi$ could be replaced by the
correlation length in describing the approach of the point
$\Delta=-1$ in \cite{Ercolessi:2010eb}. The analysis of the limit
$\Delta\to-1$, and the results  of \cite{Ercolessi:2010eb},
naturally beg the question as to the entanglement entropies of the
infinitely-many ground states of the XXZ model at $\Delta=-1$. As
we said, it turns out that the state $\psi_-$ is just one of these
ground states, although a very particular one, highly entangled.
Likewise, it is possible to construct a ground state for every
possible finite asymptotic value of the entanglement entropy and
it may be that behaviours observed in \cite{Ercolessi:2010eb}
result from approaching these various ground states. We will come
back to the study of the entanglement entropies of $\Delta=-1$
ground states in a later work \cite{twistagain}.

\paragraph{Acknowledgments:}

We are grateful to Robert Weston and Jean-Michel Maillet for
useful correspondence at an early stage of this work. We are also
indebted to Francesco Ravanini for interesting discussions on
entanglement entropy during his recent visit to City University
London and King's College London.

\end{document}